\documentclass[preprint2]{aastex}
\def\mathstacksym#1#2#3#4#5{\def#1{\mathrel{\hbox to 0pt{\lower 
    #5\hbox{#3}\hss} \raise #4\hbox{#2}}}}

\mathstacksym\lta{$<$}{$\sim$}{1.5pt}{3.5pt} 
\mathstacksym\gta{$>$}{$\sim$}{1.5pt}{3.5pt} 
\mathstacksym\lrarrow{$\leftarrow$}{$\rightarrow$}{2pt}{1pt} 
\mathstacksym\lessgreat{$>$}{$<$}{3pt}{3pt} 

\newcommand{\pref}{\protect\ref} \newcommand{\solrad}{\ifmmode{R}_{\rm
S}\else${R}_{\rm S}$\fi} \newcommand{\solmas}{\ifmmode{M}_{\rm
S}\else${M}_{\rm S}$\fi}

\newcommand{\ctn}{\ifmmode\kappa\else$\kappa$\fi}

\newcommand{\velu}{$\,$km$\,$s$^{-1}$}

\newcommand{\term}[2]{\mbox{$\,^{#1}{\rm #2}$}}
\def\term#1 #2/{\mbox{$\,^{#1}{\rm #2}$}}

\slugcomment{}

\newcommand\tabone{
\protect\begin{deluxetable}{lllll}
\protect\label{tab:tabone}
\renewcommand{\baselinestretch}{1.2}
\tablecaption{The evolution of spicule and fibril spectral imaging observations}
\tablehead{Reference& Line(s)[$n\lambda$] &  ${\cal R} = \lambda /  \Delta \lambda$
& Spatial  & Cadence\\
  & &  &  resolution & }
\startdata
\cite{Beckers1968} & H$\alpha$ / various& $2000 -20\,000$ & $\approx1-2\arcsec$
& $\gta 4$ sec\\ 
\cite{Rutten2006} & \ion{Ca}{2} 396.8[1] & 2800 & 0\farcs2 & $\approx 10$ sec\\
\cite{dePontieu+others2007} &   \ion{Ca}{2} 396.8[1] & 1800 & $0\farcs16$
&$\ge 4$ sec\\
\cite{Judge+Reardon+Cauzzi2012} & H$\alpha$[1] & $250\,000$ & $0\farcs2$ & 1 sec\\
\cite{Sekse+others2012} & \ion{Ca}{2} 854.2[16] H$\alpha$[28] &
100\,000 & 0\farcs18, 0\farcs14 & 12 sec\\
\cite{Sekse+Rouppe+dePontieu2013} & H$\alpha$[4-35] & 100\,000 & 0\farcs14 & 0.9-8 sec \\
\cite{Sekse+others2013} & \ion{Ca}{2} 854.2, H$\alpha$ & 100\,000 & 0\farcs18,0\farcs14 & various \\
This paper & H$\alpha$[1], \ion{Ca}{2} 854.2[1] & $250\,000$ & $0\farcs17,0\farcs22$ &1, 5 sec \\
                  & H$\alpha$[13], \ion{Ca}{2} 854.2[29] & $250\,000$ & $>0\farcs2$ &15 sec \\
\enddata 
\tablecomments{}
\end{deluxetable}
}

\newcommand\figone{
\begin{figure}[!ht] 
\epsscale{1.}
\plotone{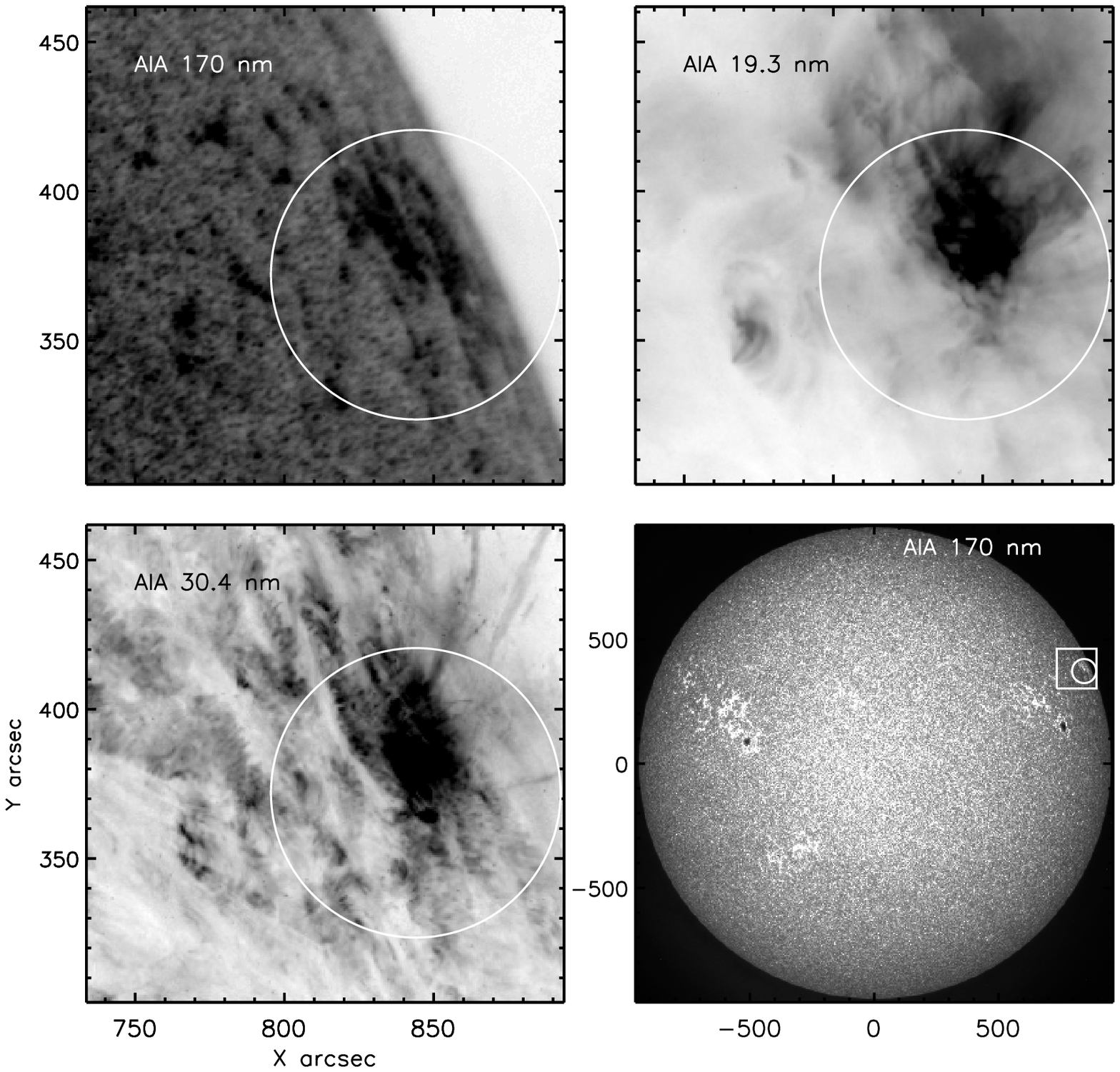}
\caption{\label{fig:context}  The  field of view of the IBIS
  observations is shown as a circle against images 
obtained between 14:25:19 and 14:25:31 UT on 7th August 2010, from the AIA instrument
  on the SDO spacecraft.  The image densities are reversed and are proportional to
  the square root of the brightness, except for the bottom right
  image. 
}
\end{figure}
}

\newcommand\figtwo{
\begin{figure}[!ht] 
\epsscale{1.}
\plotone{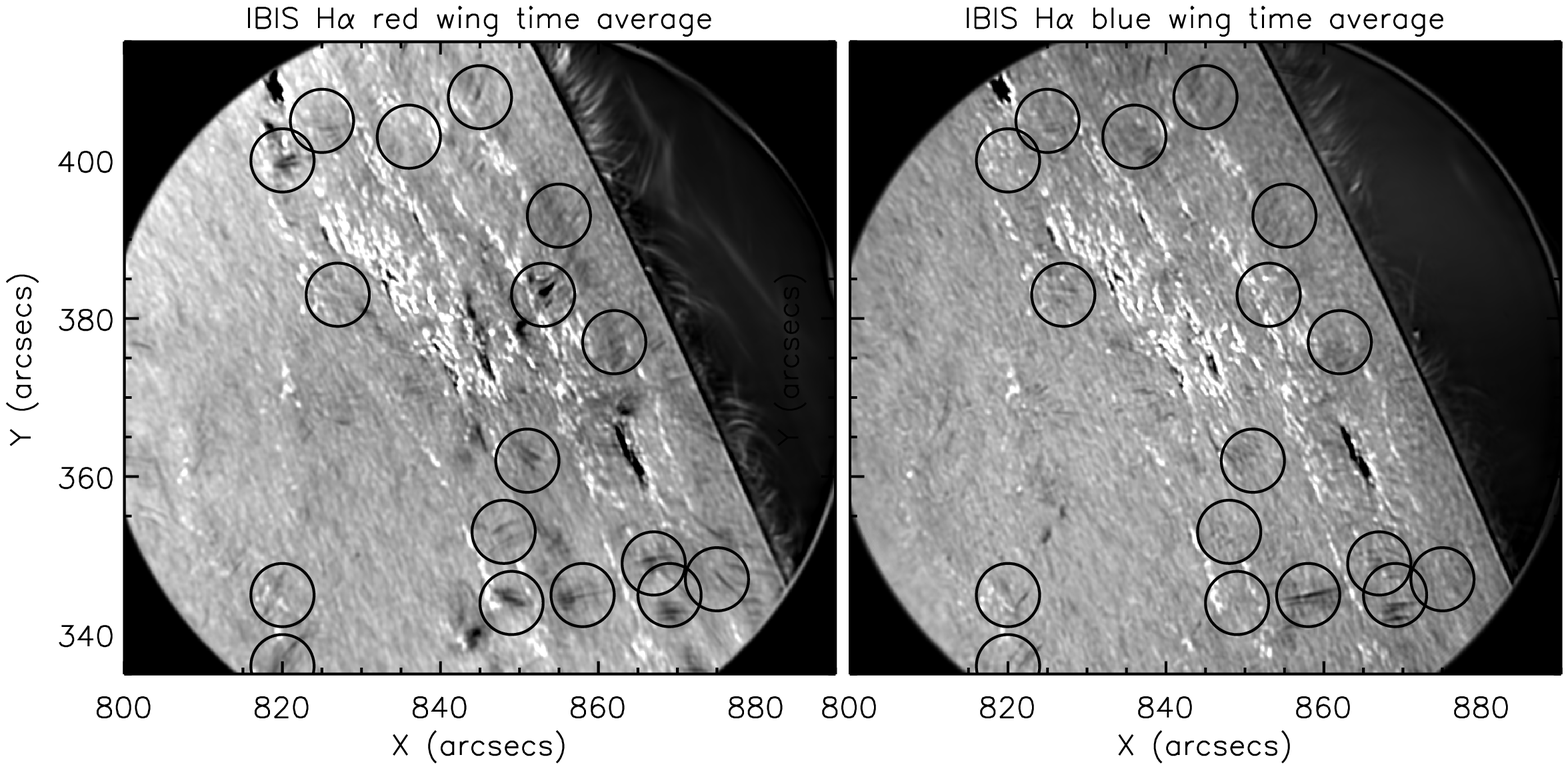}
\caption{\label{fig:average} The time average of images obtained 0.11
  nm on the red (left) and blue (right) sides of H$\alpha$ line center, from IBIS.  292
  red images
   obtained between 14:11 and 14:34 UT, and 
90
blue   images were  obtained between 14:15 and 14:23 UT, 
  each reconstructed using multi-frame blind deconvolution, were used
  to construct this figure.  A linear color table was used between the
  5\% and 95\% levels in the distributions of brightness. 
  ``Blurred'' features are rapidly
  moving chromospheric
 fine structures, including some analyzed by JRC12. 
 Photospheric features (sunspots, plages) are much
 sharper.  The circles highlight regions where properties of fibrils 
are examined in this paper.  These encircled regions appear in other figures.}
\end{figure}
}

\newcommand\figthree{
\begin{figure}[!ht] 
\epsscale{1.1}
\plotone{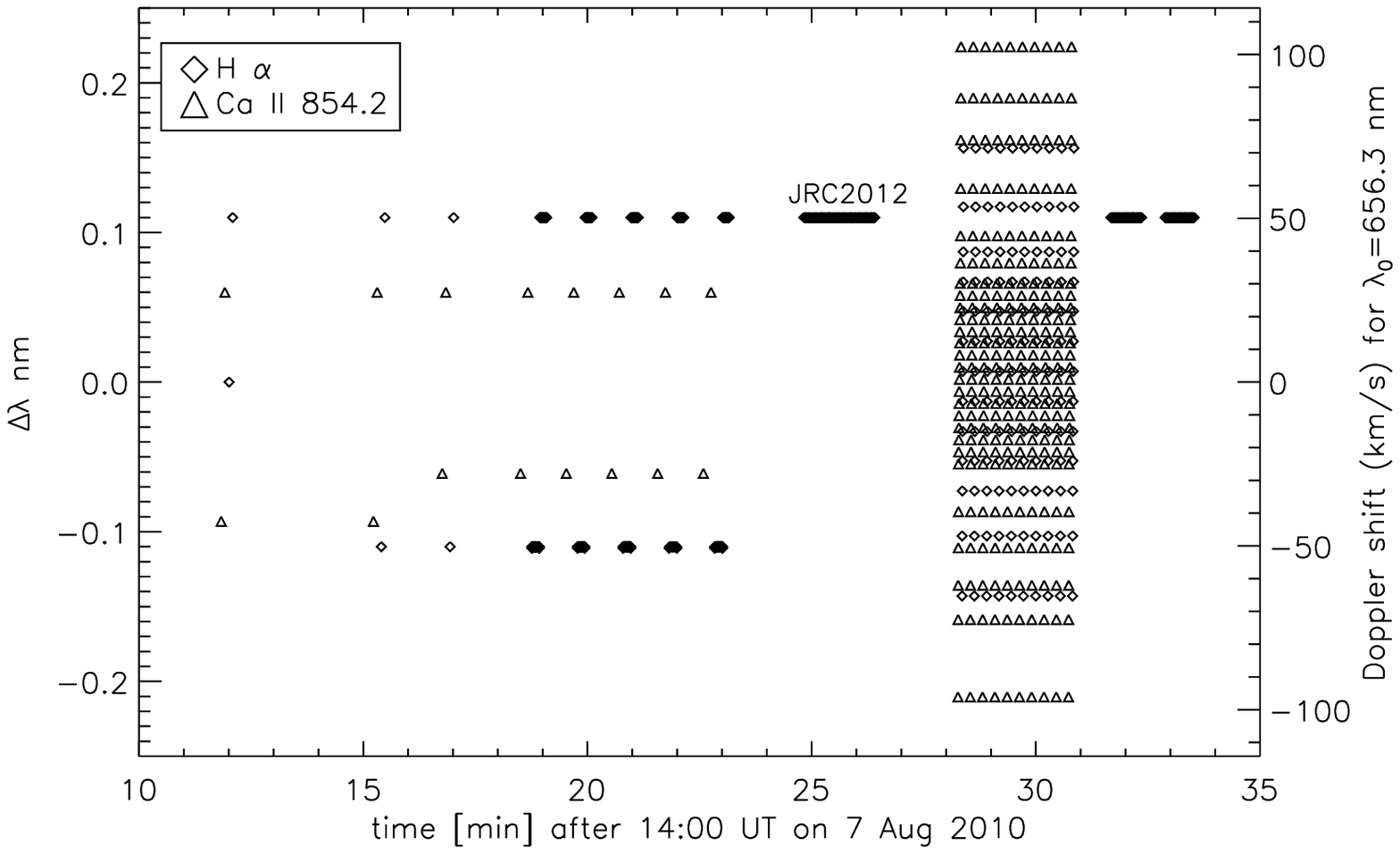}
\caption{\label{fig:timeplot} A figure showing the final set of fully
  processed intensity data acquired with the IBIS instrument on 7th
  August 2010.  The times and wavelengths of the ``monochromatic''
  images (with a spectral resolution of 200,000) in lines of H$\alpha$
  and of \ion{Ca}{2} at 854.2\,nm are shown as a function of time.
  The data marked ``JRC12'' is the 100-frame 1\,sec cadence dataset
  analyzed by \cite{Judge+Reardon+Cauzzi2012}.}
\end{figure}
}

\newcommand\figw{
\begin{figure}[!ht] 
\epsscale{.8}
\plotone{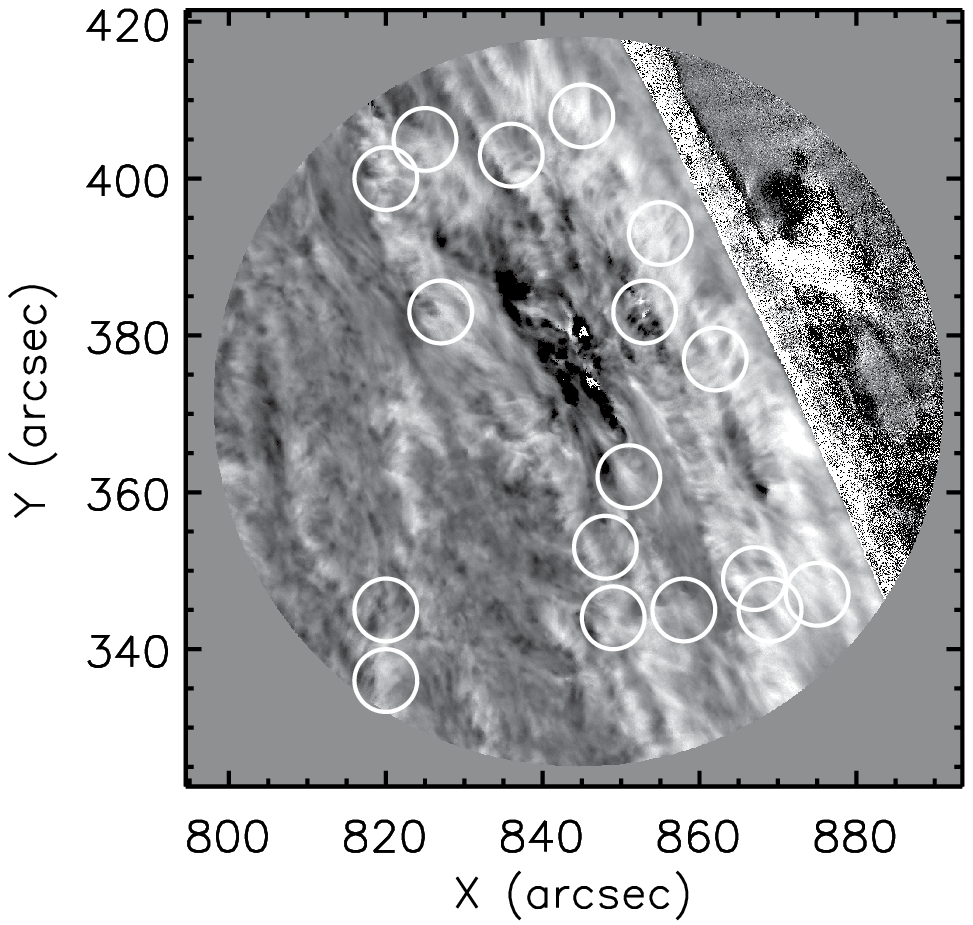}
\caption{\label{fig:widths}  The image shows line widths of
the   \ion{Ca}{2} 854.2 nm line
  computed from the first of the wavelength scans.   The color table
  is clipped at 0.04 and 0.08 nm.  This image should be compared in
  particular with the time averaged data of
  Figure~\pref{fig:average}. 
}
\end{figure}
}

\newcommand\figloops{
\begin{figure}[!ht] 
\epsscale{1.}
\plotone{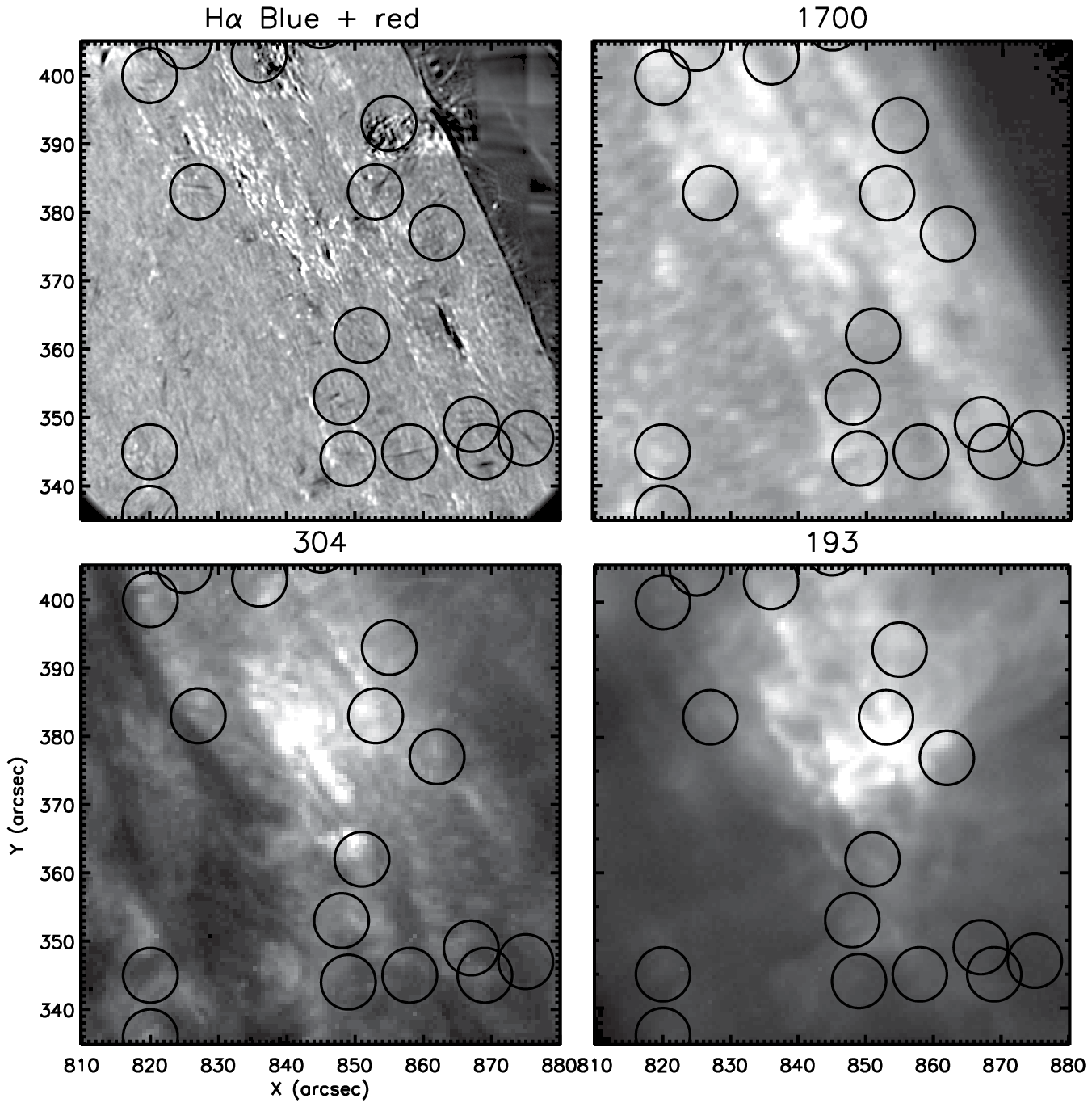}
\caption{\label{fig:loops}  A comparison of IBIS fibrils (groups are 
encircled, taken from Figure~\pref{fig:average}) 
  with data from the AIA instrument on SDO obtained between 14:25:19
  and 14:25:31 UT on August 7 2010.   
}
\end{figure}
}

\newcommand\figpcah{
\begin{figure}[!ht] 
\epsscale{.7}
\plotone{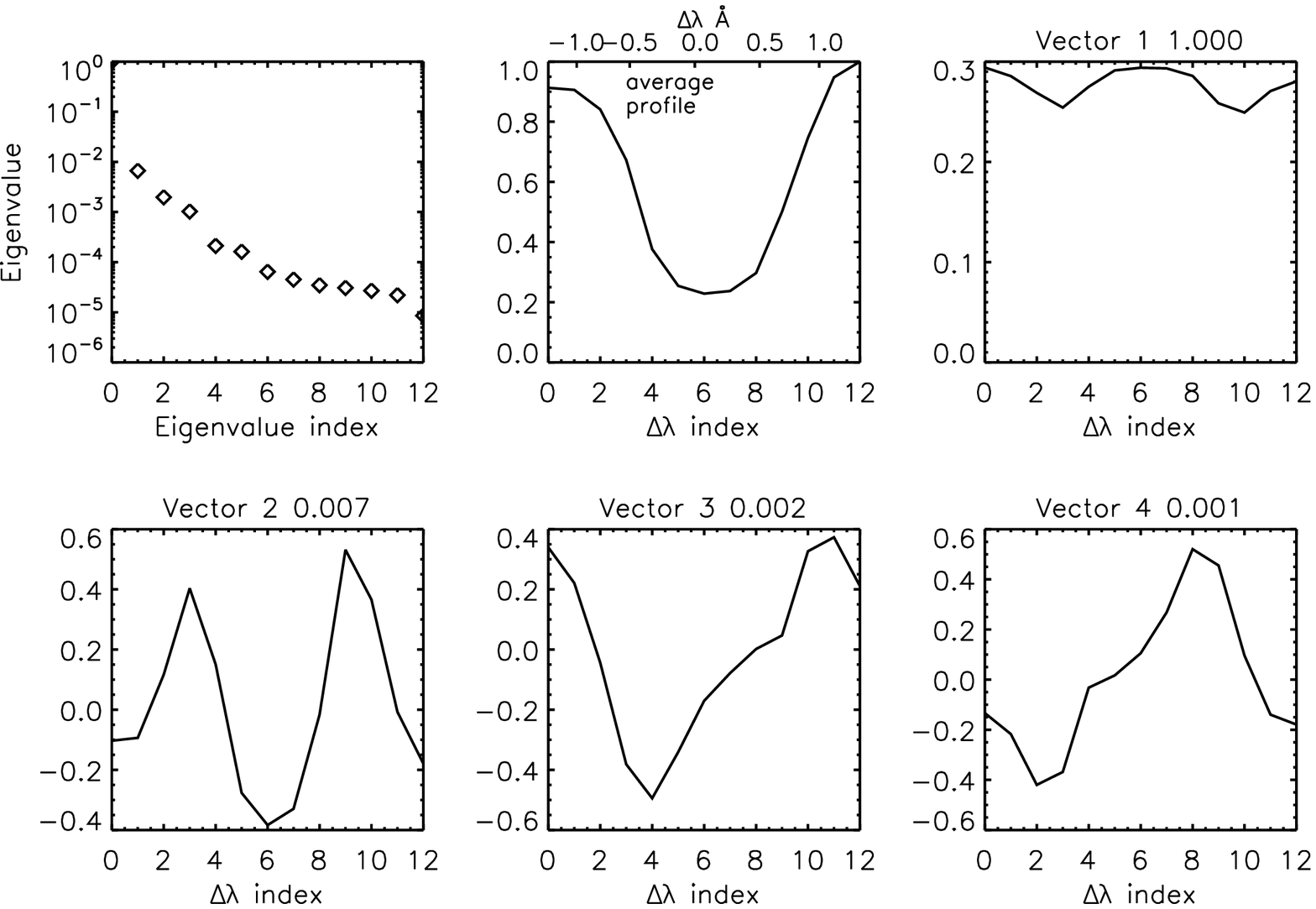}
\caption{\label{fig:pcah}
PCA decomposition of the fibril spectra in the H$\alpha$ line.  The top left panel shows the distribution of
eigenvalues, the mid top panel shows a normalized average profile for the fibrils, and 
the others show the eigenvectors that belong to the principal components.  The wavelength scales 
are shown in the average profile panel.   The y axis scales on the eigenvector components are linear but 
arbitrary, it is only the 
form of the profiles that is important. 
}
\end{figure}
}

\newcommand\figpcaca{
\begin{figure}[!ht] 
\epsscale{.7}
\plotone{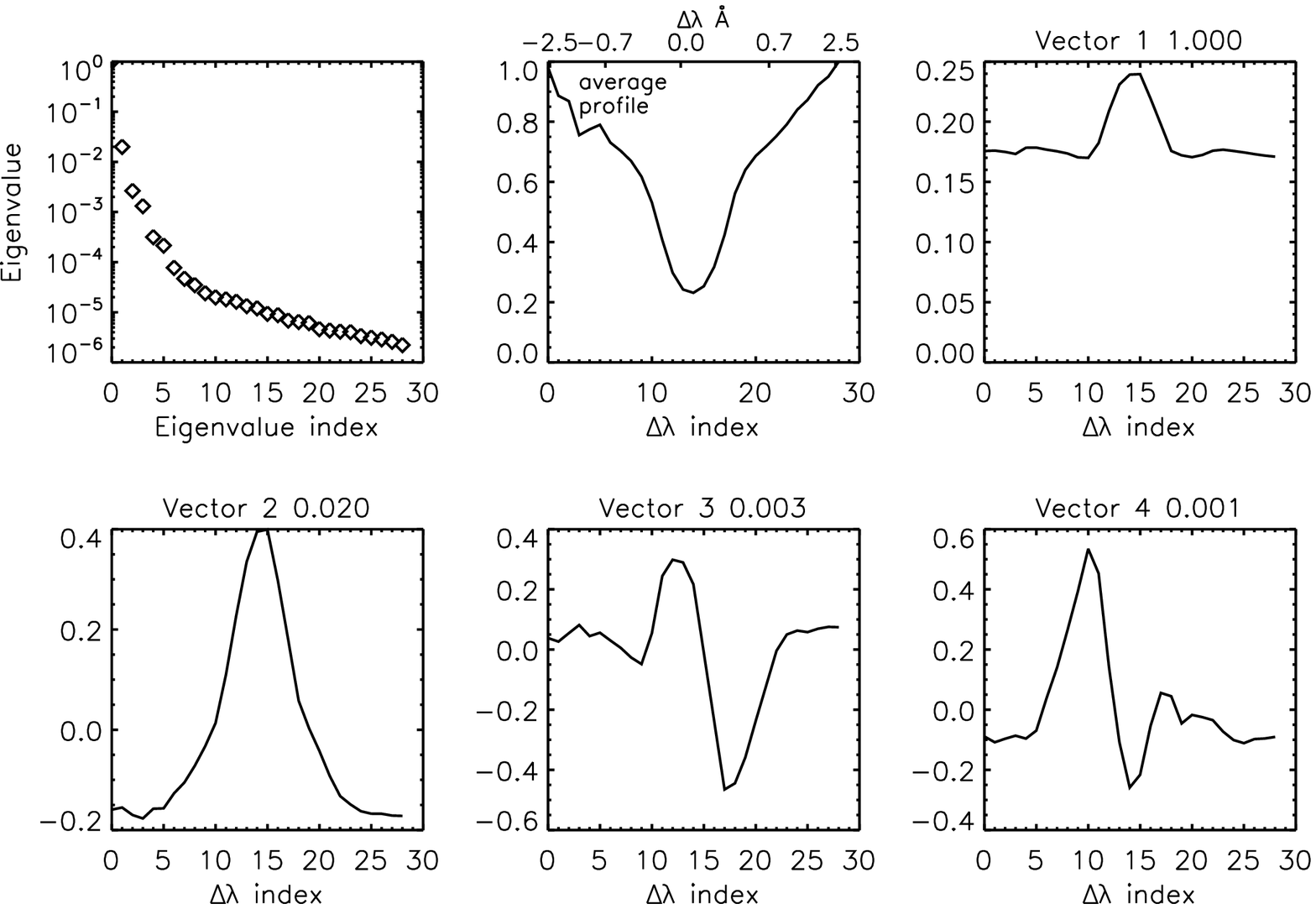}
\caption{\label{fig:pcaca}
PCA decomposition of the fibril spectra in the \ion{Ca}{2} 854.2 nm  line.
}
\end{figure}
}

\newcommand\figrvb{
\begin{figure}[!ht] 
\epsscale{.9}
\plotone{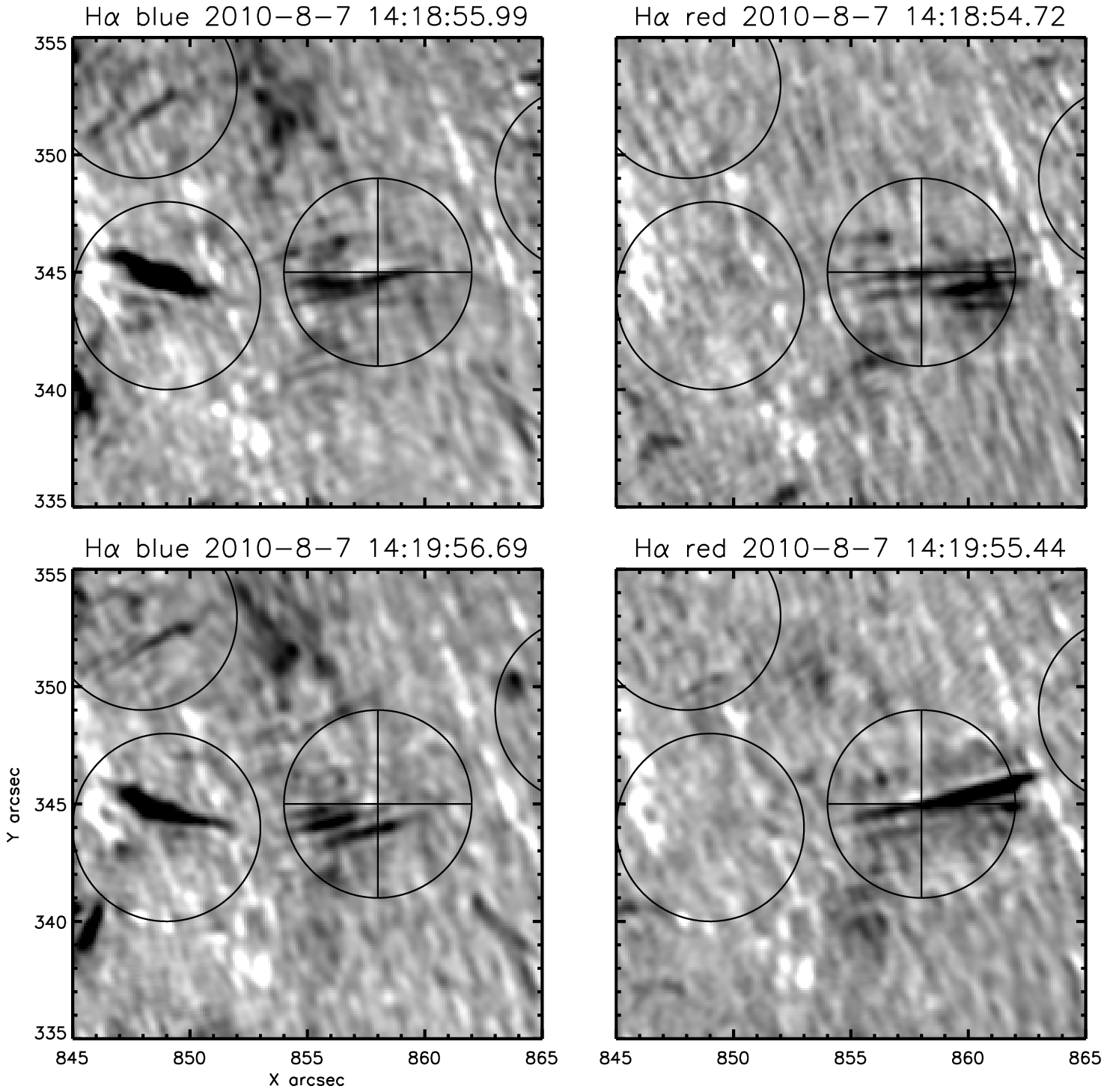}
\caption{\label{fig:rvb}  A close-up view of 
near-simultaneous H$\alpha$ blue and red wing images.($\pm0.11$ nm).
Each row contains two images obtained $\approx 1.3$ seconds apart.  The
lower images were taken 61 seconds after the upper images.
}
\end{figure}
}

\newcommand\figprofiles{
\begin{figure}[!ht] 
\epsscale{1.}
\plotone{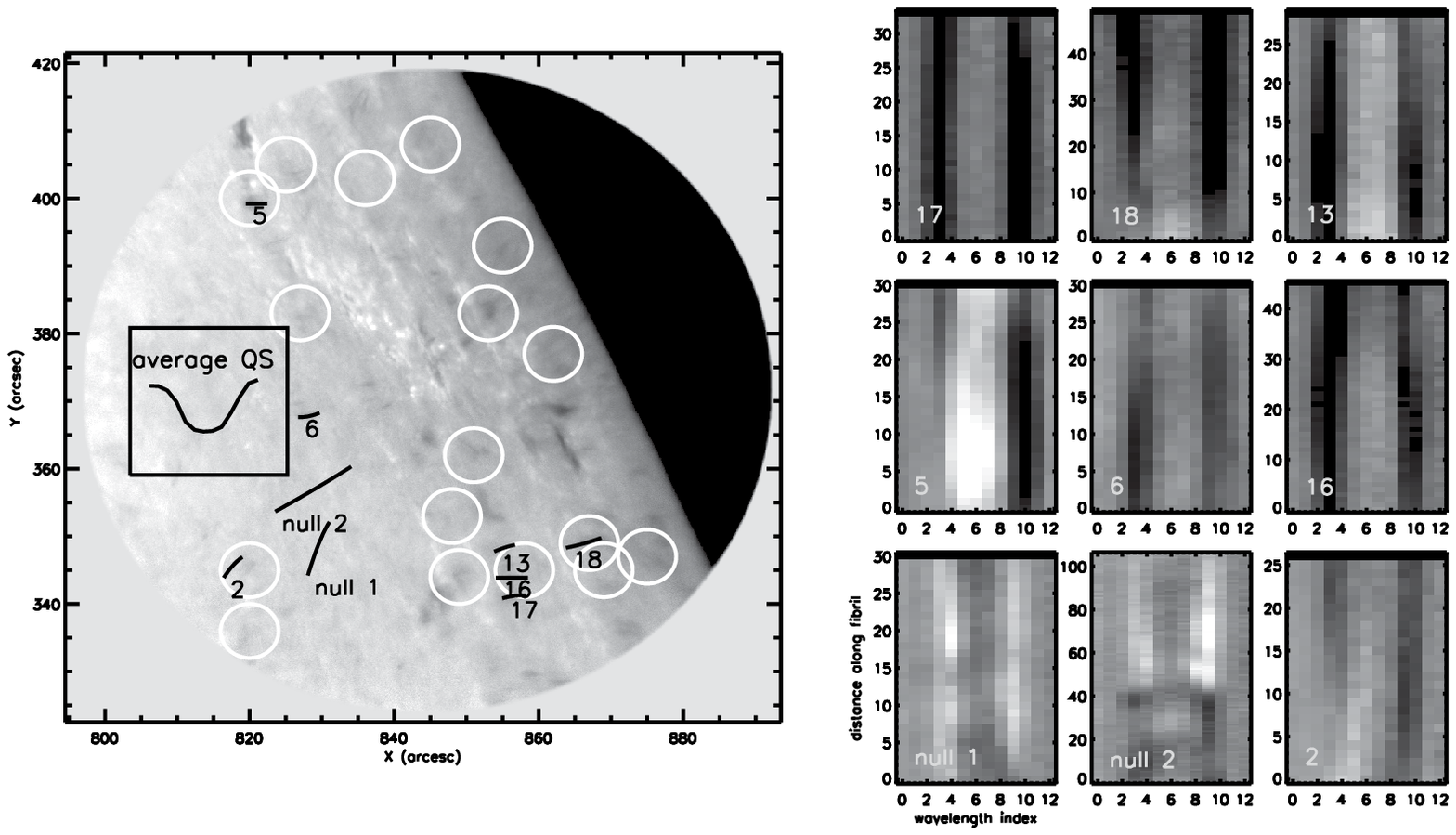}
\caption{\label{fig:profiles}  Left panel: an image obtained at +0.17
  nm from line center in H$\alpha$ annotated with fibrils and
  non-fibril regions for comparison.  The square shows an area of
  quiet Sun used to compute an average profile in H$\alpha$.  The
  smaller panels show typical profiles (divided by the average
  profile) of the regions marked in the left panel, as a function of
  wavelength index and position along the fibril.  The images have
  been scaled so that black corresponds to 0.7, white to 1.4 times the
  mean quiet Sun spectrum.  
}
\end{figure}
}

\newcommand\figprofilesc{
\begin{figure}[!ht] 
\epsscale{1.}
\plotone{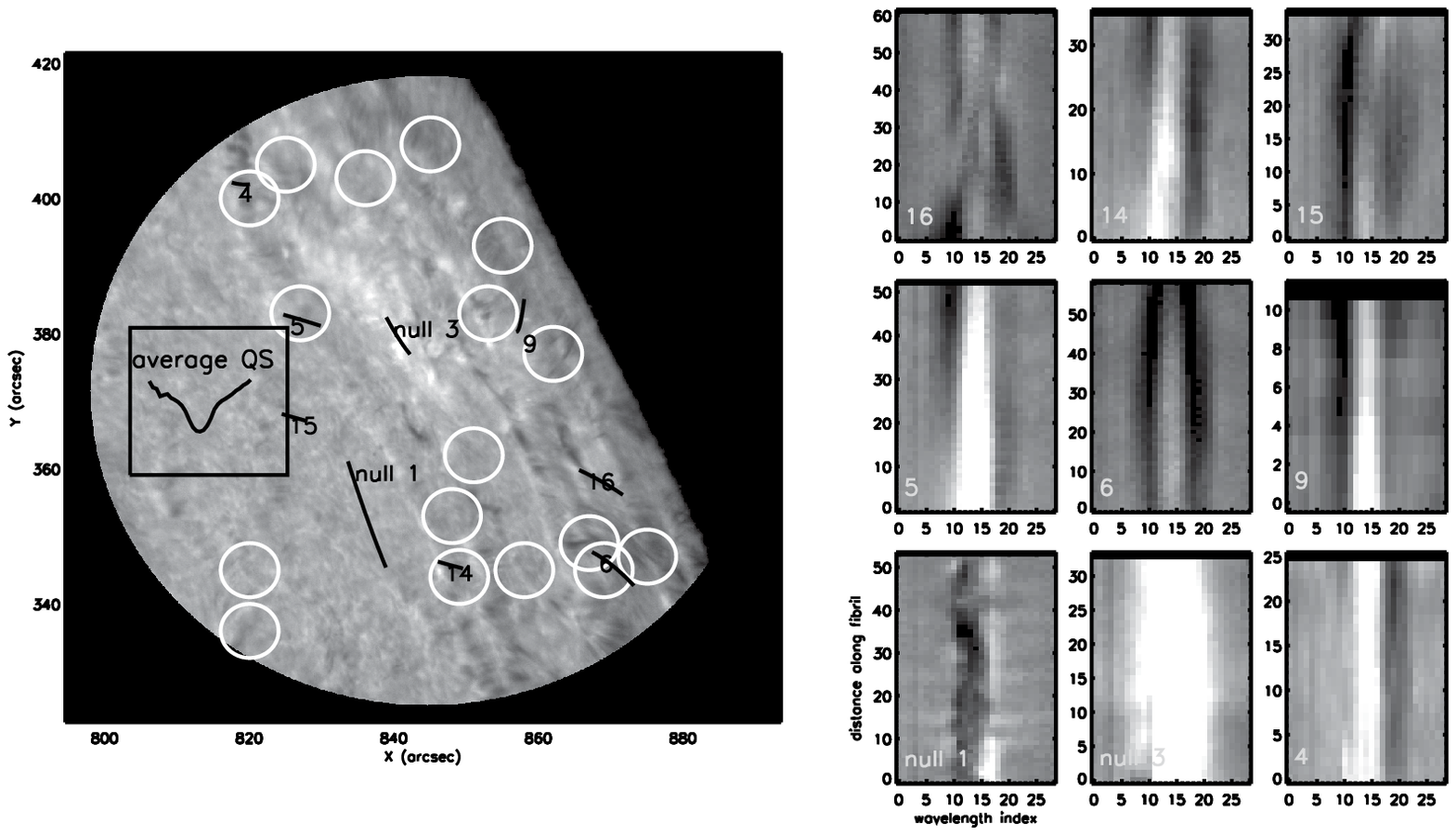}
\caption{\label{fig:profilesc}  An identical plot to
  Figure~\pref{fig:profiles}, but for the \ion{Ca}{2} 854.2 nm
  line.  Fibril number 16 appears more complex- this might be due to
  superposition of two fibrils along the line of sight.  
}
\end{figure}
}

\newcommand\figmvs{
\begin{figure}[!ht] 
\epsscale{1.}
\plotone{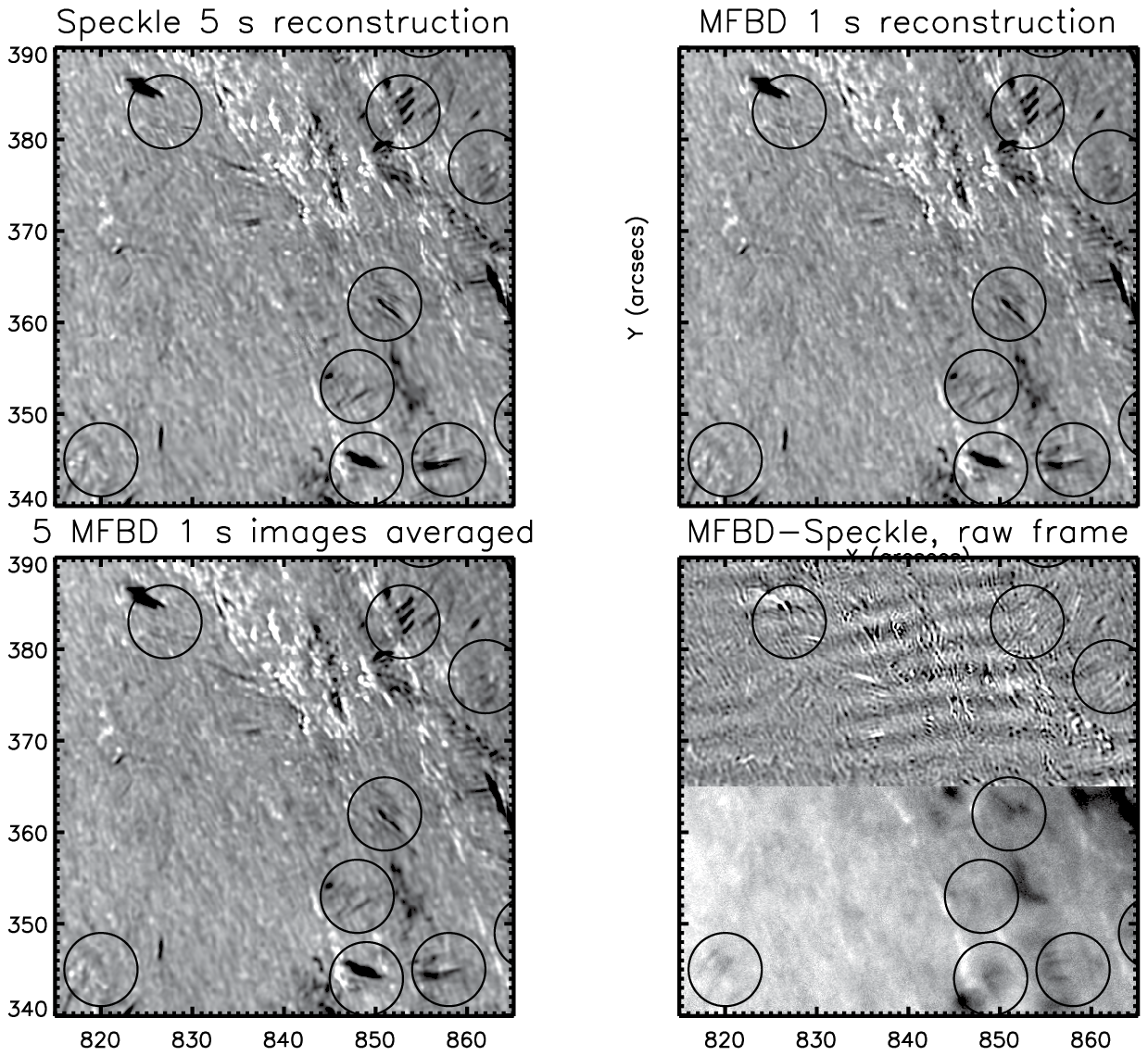}
\caption{\label{fig:mfbd_vs_spec}
  A comparison of images processed using the MFBD and speckle
  techniques at H$\alpha$ +0.11 nm.   The three complete images show data with a linear
  color table between 1\% and 99\% of the brightness distributions  The lower right image shows 
the difference between the two 5s images (upper half) scaled between
0.8 and 1.2 of the median brightness, and an 
unprocessed image from the wavelength scan
series for comparison, again scaled between 1\% and 99\%.  The
center-to-limb variation is visible in the unprocessed data.
}
\end{figure}
}

\newcommand\figacc{
\begin{figure}[!ht] 
\epsscale{1.}
\plotone{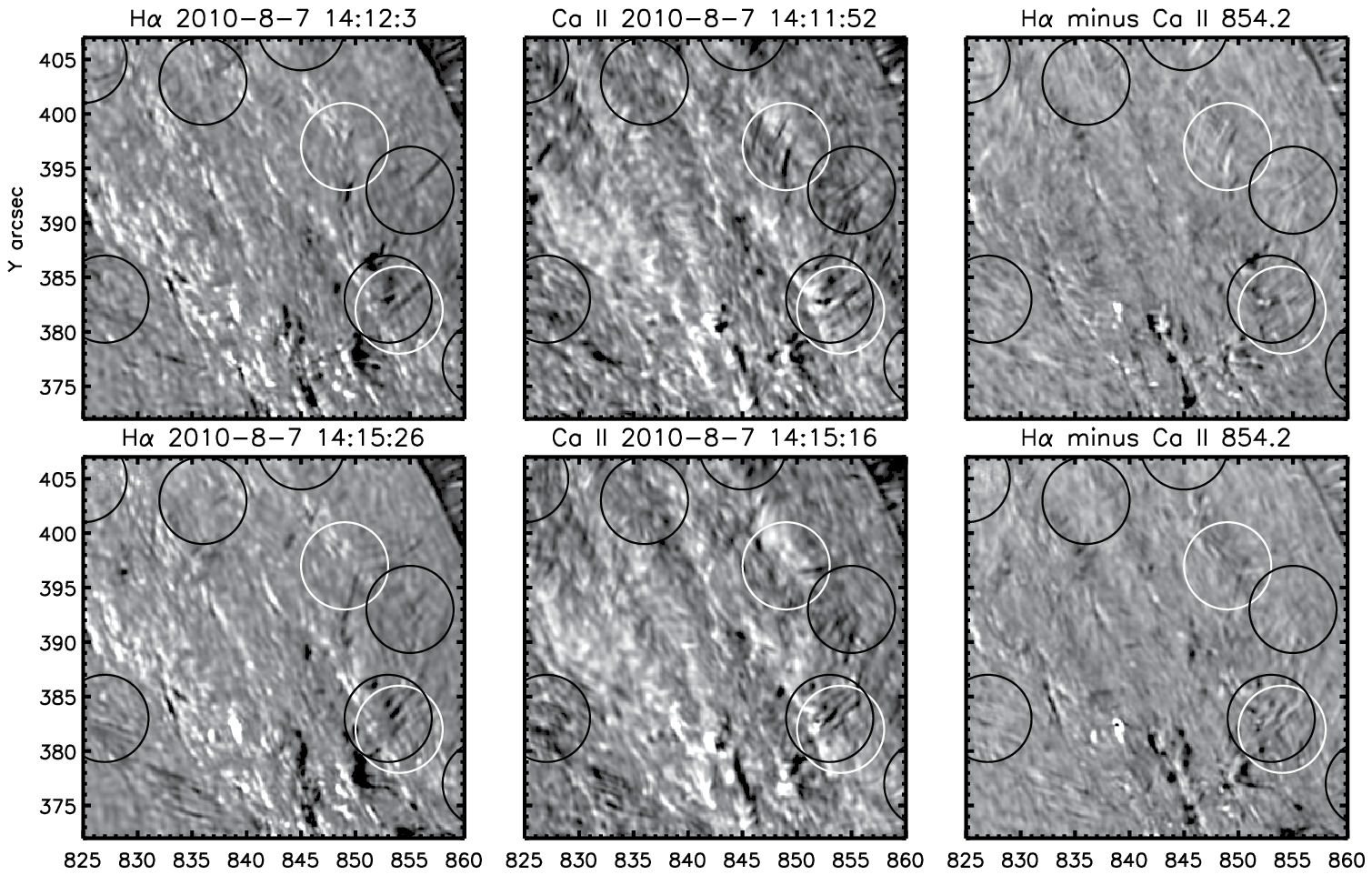}
\caption{\label{fig:acceleration}  
 H$\alpha$ images at +0.11 nm and 
\ion{Ca}{2} images at +0.06 nm from line center are shown in the
left and  middle 
panels respectively.    The right panels show 
images in which
the \ion{Ca}{2}  images were subtracted from H$\alpha$ images
obtained just 10 sec afterwards.  
   The white circles highlight regions where (upper right panel)
fibrils appear to change from white to black along its
length, indicating a continuous structure with systematically
changing thermodynamic
properties (line width, shift, depth) along it. In fact, close
inspection reveals that this is probably a chance line-of-sight
occurrence.  There are very few, if any, clear cases of such continuous behavior in
these data. 
}
\end{figure}
}

\newcommand{\philemail}{judge@ucar.edu}
\newcommand{\kevinemail}{kreardon@arcetri.astro.it}
\newcommand{\giannaemail}{gcauzzi@arcetri.astro.it}
\newcommand{\isabelemail}{iliparti@smith.edu}

\begin{document}


\title{\large The solar chromosphere observed at 1 Hz and $0\farcs2$ resolution}

\author{Isabel Lipartito} \affil{Smith College, 99 Paradise Road,
  Northampton, MA 01063, USA; \isabelemail}

\author{Philip G. Judge} \affil{High Altitude Observatory, National
Center for Atmospheric Research\altaffilmark{1}, P.O. Box 3000,
Boulder CO~80307-3000, USA; \philemail}

\altaffiltext{1}{The National Center for Atmospheric Research is
sponsored by the National Science Foundation}

\author{Kevin Reardon}
\affil{National Solar Observatory/Sacramento Peak\footnote{Operated by the %
       Association of Universities for Research in Astronomy, Inc. (AURA), %
       for the National Science Foundation}, P.O.~Box 62, 
       Sunspot, NM-88349, U.S.A.;  \kevinemail}

\author{Gianna Cauzzi}
\affil{INAF -- Ossevatorio Astrofisico di Arcetri, I-50125 Firenze, Italy; 
        \giannaemail}

\begin{abstract}
  We recently reported extremely rapid changes in chromospheric fine
  structure observed using the IBIS instrument in the red wing of
  H$\alpha$.  Here, we examine data obtained during the same observing
  run (August 7 2010), of a mature active region NOAA 11094.  We
  analyze more IBIS data including wavelength scans and data
  from the Solar Dynamics Observatory, all from within a 30 minute
  interval.  Using a slab radiative transfer model, we investigate the
  physical nature of fibrils in terms of tube-like vs. sheet-like
  structures.  Principal Component Analysis shows that the very rapid
  H$\alpha$ variations in the line wings depend mostly on changes of
  line width and line shift, but for \ion{Ca}{2} 854.2 the variations
  are dominated by changes in column densities.  The tube model must
  be rejected for a small but significant class of fibrils undergoing
  very rapid changes.  If our wing data arise from the same structures
  leading to ``type II spicules'', our analysis calls into question
  much recent work.  Instead the data do not reject the hypothesis
  that some fibrils are optical superpositions of plasma collected
  into sheets.  We review how Parker's theory of tangential
  discontinuities naturally leads to plasma collecting into sheets,
  and show that the sheet picture is falsifiable.  Chromospheric fine
  structures seem to be populated by {\em both} tubes {\em and}
  sheets.  We assess the merits of spectral imaging versus slit
  spectroscopy for future studies.
\end{abstract}


\section{Introduction} \label{sec:introduction}

We have previously questioned the traditional  interpretation that chromospheric
fine structure, as revealed by H$\alpha$ data for example,  
corresponds to plasma that fills narrow tubes of
magnetic flux \citep[][henceforth JTL11, JRC12]{Judge+Tritschler+Low2011,
  Judge+Reardon+Cauzzi2012}.  In JRC12 we argued that at least some newly
observed properties of the fine structure are inconsistent with the
traditional picture, and instead have proposed that the fine structure
contains at least some sheet-like structures, corresponding perhaps to
the magnetic tangential discontinuities (TDs) whose existence has been
argued on theoretical grounds by \citet{Parker1994}.  The primary
observational evidence we cite is the rapid appearance and
disappearance of ``typical'' fine structures simultaneously along
their entire length of 3--5,000 km, 
seen in absorption at one particular wavelength
in the H$\alpha$ line.  By ``rapid'' we mean super-Alfv\'enic apparent
speeds, so that no propagating disturbance can account for the rapid
appearance of a feature all along its length.  The observed behavior 
led us to discount oblique wave propagation and standing
waves, the latter on the grounds that the features appear suddenly out of nowhere, whereas
standing waves require special boundary conditions and 
finite times to become established.   This point stands in 
contrast to the claims of others, and it is addressed here in our  Appendix A.

In the light of the recent launch of the IRIS spacecraft, and 
to shed further light on the problem, here we re-analyze data reported
earlier (JRC12) with additional data. 
We try to refute  two competing hypotheses: (1) that fine
structure corresponds to plasma constrained to flow along and within
narrow tubes or ``straws'' of magnetic flux, and (2) that fine
structure corresponds to material trapped within surfaces of
tangential discontinuity embedded in an otherwise continuous
structure.  The answers are not just of academic interest, important
physical consequences include the following: the interpretation of apparent motions
in terms of real motion vs. phase effects; the existence of magnetic
TDs in plasmas in a low $\beta$ state ($\beta$
is the ratio of plasma to magnetic pressure); the long standing
problem of energy transport in the lower transition region.

We regard both hypotheses as worthy of study.  Indeed,
the evidence in support of tubes is strong in photospheric features
such as sunspot penumbrae, and in certain kinds of chromospheric 
phenomena called ``dynamic fibrils''  \citep{Hansteen+others2006},
later associated with the 
``spicule of type I''  documented by \cite{dePontieu+others2007}.  But we believe that 
the question remains more open for the very
different spicules of type II  that are believed to carry mass,
momentum and energy into the corona (see, e.g., \citealp{Zhang+others2012}),
despite claims 
that the evidence only supports the traditional,
flux tube picture  \citep[e.g.][]{Sekse+Rouppe+dePontieu2013}. 
Several studies based upon further theoretical and observational 
considerations 
have been
published questioning the role of type II spicules  in coronal physics 
\citep{Klimchuk2012,Tripathi+Klimchuk2013,Patsourakos+Klimchuk+Young2013}.

 Below we refer to all
dark ``chromospheric fine structures'' which appear on the solar disk
like thin, straw-like structures, simply as ``fibrils''.  


\section{Observations}

Spectroscopic observations of a limb region, covering a 97\arcsec{}
diameter circular field, were obtained on 7th August 2010 using the
IBIS instrument
\citep{Cavallini2006} with the Dunn Solar Telescope at Sunspot, NM. 
The field of view centered near the NW solar limb
between 14:10 and 14:35 UT.  Co-alignment of the time average of
images obtained in the red wing of H$\alpha$ with plages and sunspots
in an image at 170 nm obtained at 14:25 UT from SDO AIA yields
a target center of
$x=844\arcsec$, $y=372\arcsec$, with an uncertainty of an arc second.
Images of the low chromosphere, transition region and corona  from the AIA instrument
\citep{Lemen+others2012} on the Solar Dynamics Observatory (SDO)
spacecraft includes the field of view of the IBIS observations in
Figure~\pref{fig:context}.  The field of view observed includes a
mature active region (NOAA 11261) with a small sunspot which was used
as the lock point for the adaptive optics system. This region produced
many C- and several M- class flares during the preceding disk passage,
but flaring had stopped two days before our observations were made.  

Time averages of red and blue wing IBIS images are shown 
in Figure~\pref{fig:average}.  The circled areas highlight regions with
significant numbers of narrow, long fibrils; they are seen in this
average plot as ``smudges'' as they come and go during the averaging.   
The scans run with IBIS are summarized in Figure~\pref{fig:timeplot}.
Each point in the plot represents one fully processed image (see
below) at a given
wavelength.  There are several time series shown (many closely packed
abscissa points at one ordinate position), as well as some full wavelength
scans between 14:28 and 14:31 UT.  The longest time series, marked
``JRC2012'', was analyzed by JRC12.

The raw IBIS data consist of counts as a function of
position on the sky, the region observed being shown in  Figure~\pref{fig:average}.  Each exposure is of 60 msec 
integration time except for two H$\alpha$ bursts at 14:31:32 and
14:32:46 (500 frames each) which had 25 msec exposures.  
The camera accumulated data at a 0.1s cadence.  The data were processed to include standard
dark, flat and linearization corrections.  The time series data -- all
those data shown in Figure \pref{fig:timeplot} except the wavelength
scans obtained between 14:28 and 14:31 UT -- were further processed
using speckle interferometric \citep{Woeger+vonderLuhe+Reardon2008}
and multi-frame blind deconvolution (MFBD) techniques
\citep{Lofdahl2002}, to correct residual seeing-induced errors.  The
cadence of the seeing-corrected time series datasets is 1 sec (MFBD) 
and 5 sec (speckle).  In the time-series reconstructions  center-to-limb brightness
variations are removed. The
wavelength scans also included corrections for the slightly variable
wavelengths across the field of view that results from the ``classical
mount'' of the IBIS etalons.  Such corrections, relying on
interpolations in wavelength, cannot be applied to
the fixed-wavelength, time series data,
but its effects are irrelevant to our results as we deal mostly with relative variations at any given spatial
position.   


The MFBD and speckle image reconstructions are not
identical, as expected.  
Figure~\pref{fig:mfbd_vs_spec} compares MFBD and speckle images taken
at $+0.11$nm from line center, over a small region of the FOV.  
We built the MFBD
dataset from ten consecutive 0.1s frames, but the speckle dataset was
built from fifty frames.   
In the Figure 
we juxtapose two sets of
speckle and MFBD images taken in H$\alpha$. These reconstructions are built from the same 
raw frames, highlighting 
the differences
between the two techniques, as well as differences between data recovered
at 1 and 5 second cadences.
While there is an observable difference between
MFBD and speckle images in terms of definition of structures,
fringe-like artifacts and 
brightness/depth levels, the differences are small compared with the
robust properties of the obvious fibril-like features analyzed here.

The figure also shows (lower right panel, bottom half of the image) 
an unprocessed image from the wavelength scan
series.  The unprocessed image is of much lower resolution,
showing the power of reconstructions.  This difference must be kept in
mind
when comparing reconstructed data with the wavelength scan data
below (section \pref{sec:scans}). 
Clearly,  post-processing as well as adaptive optics are important for
the measurement of such fine structures as fibrils.

\section{Analysis}

\subsection{A simple model}
\label{sec:model}

We adopt a 
simple model for the formation of the fibril intensities. 
We consider the fibril absorption to occur in  a  slab of plasma
which has associated with it microscopic (thermal and turbulent)
velocities, 
and macroscopic line-of-sight (flow) 
velocities.  These quantities translate to  a local Doppler line width $w$ and shift $s$
respectively.   Our goal is to relate changes in the line intensity
at a given frequency to changes in the minimum number of parameters
required: these are the line
optical depth, the line shift, and the line width. 
Assuming the source function in the slab is small compared
with the photospheric intensity $I^0_x$, and slab optical depths $\delta\tau_x$ are
$\ll 1$, the change in intensity at a  (normalized) frequency $x$
through the slab, 
is (e.g. JRC12)
\begin{equation}
  \label{eq:1}
    \delta I_x = -I^0_x \delta\tau_x\, = -I^0_x \, a_{ij} \, \times
    \end{equation}
$$
    \left  (  \phi_x \delta (n_i \Delta)  +
(n_i \Delta) \left \{  
  \left ( \frac{\partial \phi_x}{\partial s} \right )_w {\delta s} +
  \left ( \frac{\partial \phi_x}{\partial w} \right )_s {\delta w}
\right \}
 \right ).     
$$
Here  $a_{ij}$ is  the absorption cross section for  
radiative transitions between atomic 
levels $i$ and $j$ ($a_{ij}=\frac{\pi e^2}{mc} f_{ij}$ with $f_{ij}$ the
absorption oscillator strength); $I^0_x$ the
incident photospheric intensity; $\phi_x$ the normalized absorption
profile  ($\int_{-\infty}^{\infty} \phi_x dx=1$);  $n_i \Delta$ is the population density of the lower level of
the line transition times the geometrical thickness $\Delta$ of the absorbing slab.  
In this model changes in $(n_i\Delta)$, $s$ and
$w$ lead to changes in $I_x$ at frequency $x$.  

\newcommand{\gaus}{g_x}
Further insight is
gained by assuming that $\phi_x$ is a Gaussian function:
$$\phi_x= \gaus(s,w) = \frac{1}{w\sqrt{\pi}} \exp \left ( - \frac{(x -
    s)^2}{w^2} \right )$$  
We adopt Dopper units (\velu) for frequency $x =
(\nu-\nu_0)\frac{c}{\nu_0}$,
$c=3\times10^5$ \velu, and we drop the variable names held constant
from the 
partial derivatives, for notational clarity.
 For Gaussian profiles
$ \frac{\partial \gaus}{\partial s}  = - \frac{\partial
  \gaus}{ \partial x} $, 
$ \frac{\partial \gaus}{\partial w}  = +  \frac{w}{2} \frac{\partial^2
  \gaus}{\partial x^2} $, 
therefore, to leading orders, this equation gives
\begin{equation}
  \label{eq:2}
\delta I_x =  I^{0}_x  a_{ij}  \left ( \gaus \delta(n_i\Delta)  -
  (n_i\Delta) \left \{  \frac{\partial\gaus}{\partial x} \delta s - 
  \frac{w}{2}\frac{\partial^2\gaus}{\partial x^2}  \delta w \right\} \right ).
\end{equation}
The 
changes in monochromatic intensity at each frequency $x$ which are 
caused by small changes in the parameters 
 $(n_i\Delta)$, $s$ and
$w$ are proportional to $\gaus$, $\partial\gaus/\partial x$
and $\partial^2\gaus/\partial x^2$ respectively.  
Further, we have
\begin{eqnarray}
   \frac{\partial \gaus}{\partial s}   &=&
  2\frac{(x -s)}{w^2} \, \gaus  \label{eq:3}\\
   \frac{\partial \gaus}{\partial w}   &= &
  \frac{2(x -s)^2-w^2}{w^3} \, \gaus \label{eq:4}
\end{eqnarray}
Assuming initially that variations in intensity (equation \pref{eq:1}) arise
only from changes in $s$ and $w$, for wing intensities $|x-s| > w$,
these equations show that changes of a given magnitude in $w$ lead to
$\delta I_x$ that exceed only by a little changes from the same
magnitude change in $s$, and vice-versa for core frequencies ($|x-s| <
1$).  Thus {\em similar magnitude changes in line shifts and widths influence the
  changing wing intensities to a comparable degree. }  
Below, we will find that those line profile changes responsible for
wing fibrils appear to be dominated by
first and second derivatives of the profile.   
These fibrils are visible because a particular combination of randomly
oriented (line width) and
systematic (line shift) Doppler velocities produces
absorption at the observed wavelength.  The presence of fibrils therefore reflect
both microscopic (``thermal'') and macroscopic (``flow'') velocity
fields.  They  should not be interpreted simply in terms of evidence
for a flow at the ``observed velocity'' \citep{Rouppe+others2009}.  

We will demonstrate, {\em post-facto}, that 
H$\alpha$ fibrils are made manifest mostly from changes
in $s$ and $w$ in the underlying physical structure(s) -
(section \pref{sec:interp}).  For \ion{Ca}{2} we will however also 
find significant
contributions from changes in $n_i \Delta$.   
 We will also find that fibril intensities vary by $\approx 10$\%
from the local, non-fibril, intensities.  Using equations \pref{eq:2}
- \pref{eq:4} we can solve for values of $\delta w$ and $\delta s$
that yield such intensity changes.  For H$\alpha$, using a Gaussian
line-width parameter $w = 32$ \velu{} (a combination of thermal and
non-thermal broadening giving widths similar to average quiet Sun
profiles), a central line intensity that is 23\% of the continuum, and at 
$x = 2w$ (0.11 nm) we find that $\delta s =
10$ \velu, or $\delta w = 8$ \velu{} produce $10\%$ amplitude
intensity changes.  
So, both changes in $s$ and $w$ contribute in similar measure 
and in modest amounts to the observed 10\% variation in intensity. 
 
\subsection{Wavelength Scans}
\label{sec:scans}

To supplement our analysis of time series data, we examine first the wavelength scans
obtained between 14:28 and 14:31 UT.  The good atmospheric seeing and
adaptive optics correction allow us to draw some conclusions, even if
no image reconstruction was possible (see
Figure~\pref{fig:mfbd_vs_spec} for a typical example of a wavelength
scan image).

\newcommand{\avq}{\overline{I}_x}

We identified and traced some typical fibrils visible in the
wavelength scan data at 0.11 nm and 0.06 nm for H$\alpha$ and
\ion{Ca}{2} 854.2 nm respectively. We also selected a nearby 
region from which to derive average line profiles.  The major
axes of the fibrils were traced and fitted with a quadratic in the
solar $x,y$ frame.  Individual spectra $I_{x,p}$ were then
extracted as a function of 
$x$ and pixel position $p$
along each fibril.  Figures~\pref{fig:profiles} and
\pref{fig:profilesc} show the fibrils selected, the regions used to
calculate average spectra, $\avq$, neighboring but 
outside of fibrils
(left panels), and then the spectra $I_{x,p}/\avq$.  Both
fibril and also some non-fibril (``null'') profiles are shown in the
figures.

The fibril profiles are similar to one another yet significantly
different from those of the ``null'' features plotted for comparison.
The variations in intensity are  10-20\% from fibril to 
non-fibril regions, the figures have color tables clipped at $80\%$
and $120\%$ of $\avq$.
The fibrils typically show shallower but broader absorption lines than
the average profile, akin to the broad profiles described by
\citet{Cauzzi+others2009} around magnetic network elements. Such
profiles lead to a bright line core, darker near wings, and ``normal''
far wings clearly seen in the figure.  From
equation~\pref{eq:2}, this wavelength dependence of $I_{x,p}/\avq$, almost
symmetric around line center and similar to the second derivative of
the line profile, is readily explained by fibrils having a {\em
  broader line width} $w$.  But, of twenty such fibril profiles (not
all shown), seventeen have noticeably asymmetric profiles over
significant fractions of their lengths.  Thus not only are fibril
profiles wider but the the majority 
also have significant apparent velocity shifts. 

In Figure \pref{fig:profilesc} we show, for completeness, the profiles
for \ion{Ca}{2} fibrils, in the same fashion as for
Figure~\pref{fig:profiles}.
In the Discussion section, we examine the modest variation in these
profiles as a function of distance along the fibril in an attempt
assess if they are compatible with standing waves.  But we warn against an over-interpretation of
these profile data which, more adversely affected by residual seeing
than the time series data, must be an ill-defined spatial average of
actual solar profiles.

\subsection{Statistics of fibrils}

\label{sec:stats}

In JRC12 we presented several cases of extremely rapid changes along
fibrils, our objective here is to examine the statistical occurrence
of such behavior.  
Time averaged 
H$\alpha$ images on the red and blue sides of the line are shown in
Figure~\pref{fig:average}.  Time averages are useful in highlighting
``fuzzy'' areas in which rapid spatial variations in fibrils are
present.   These areas are related to areas of enhanced line widths
over the chromospheric network \citep{Cauzzi+others2009}. 
A line width image (from the first \ion{Ca}{2}  spectral scan) 
is shown in Figure~\pref{fig:widths}. A comparison
with Figure~\pref{fig:average} reveals that the fuzzy areas are found
within only some regions of enhanced line width. Thus a broad line
appears a necessary but not sufficient condition for the rapid
variations at 0.11 nm to be observed, consistent with the behavior of
the line profiles found above.  
We have counted the number of detectable fibrils seen against
the solar disk in sets of 40 
selected blue- and 56 red-wing MFBD processed images from our sample
of H$\alpha$ data.  
We found an average of 48 and 74
fibril-like structures per frame respectively.  The distributions of the number of
occurrences of finding $n$ fibrils per image are different.
The area of the solar disk in the images is about 100 square degrees
(in heliographic coordinates) or about 0.03 steradians, about 1/400th
of the solar surface, so at any time we might find 
$2\times 10^4$ blue- and $3\times10^4$ red-wing fibrils
respectively, at these particular wavelengths.   
The sum of these
numbers is within a factor of two of some estimates of numbers of 
type II spicules and their disk counterparts 
\citep{Sekse+others2012}. 

The data were acquired in bursts (Fig.~\pref{fig:timeplot}), which
means that fibril lifetimes, while encoded into the time series,
cannot easily be determined because of burst durations are too short.
The lifetimes of the observed structures appear to vary from a few
tens of seconds to longer than the duration of the longest burst (100
seconds).  Results discussed below (see Figure~\pref{fig:acceleration})
show that lifetimes are probably shorter than 200 seconds.

We estimate by inspection 
that the fraction of fibrils exhibiting the extremely high
phase speeds highlighted by JRC12 is between 1/3 and 1/6 (red wing
images) and about 1/8 (blue wing).  The reader can judge for
his- or herself  the estimate for the red wing data using 
online movies presented by JRC12.
These small fractions do not
undermine the essential arguments of JRC12, a single occurrence
being sufficient to lead to the dilemma discussed by them.
The majority of other fibril structures appear to evolve more
slowly. Like spicules of type I 
they have smaller geometric aspect ratios (length/width), and include
some clear examples of ``coronal rain'' (blobs of cool material 
falling out of the corona) as well as other features that appear compatible
with fluid motions along tubes of magnetic flux.

\subsection{Spatial relationships between blue- and red- wing fibrils}
\label{sec:rvb}

Our data contain several instances where a red wing image was acquired
within 2 seconds of a corresponding blue image (Figure
\pref{fig:timeplot}), for each line.  Since fibril lifetimes are at
least an order of magnitude larger, except for the relatively small
fractions (1/8-1/3, section \pref{sec:stats}) of fibrils exhibiting
the frame-to-frame changes highlighted by JRC12, data taken just 2
seconds apart give almost simultaneous views of the Sun at the two
wavelengths (for the majority of fibrils).  
Figure \pref{fig:rvb} shows a typical example of the red
and blue images for data  taken 
1.3 seconds apart, and 1 minute apart.  
Statistically, H$\alpha$ red images are 
consistently longer than those in the blue images.  
The $x-y$ positions of the red and blue fibrils almost
never overlap, even though their axes are mostly aligned because they
are rooted in clumps whose magnetic field lines appear to be part of a
larger magnetic structure \citep[e.g.][]{Athay1976}.  In those
cases when red and blue fibrils are adjacent, 
there is significant asymmetry between these
neighboring red and blue fibril shapes (see, for example,
fibrils
in the cross-hairs in the figure). 
The fibrils immediately below the cross-hairs 
are in different positions in the
red and blue images.  
Those neighboring frames at different wavelengths in the \ion{Ca}{2}
time series data show qualitatively similar results, except that 
the fibrils' lengths are not noticeably different in
the red vs blue sides of the \ion{Ca}{2} lines.  

In summary, close examination shows that red- and blue- wing fibrils
can be found that are sometimes co-spatial and often parallel.  Indeed
they must be co-spatial for the small fraction of fibrils showing
almost symmetric profiles (section \pref{sec:scans}).  However, the
majority of the spectra appear asymmetric. Consequently the majority
of fibrils are not co-spatial.  If, as some have suggested \citep{dePontieu+others2012}, torsional
motion is important, we should see evidence for this in our data instead.
Those red and blue fibrils which appear
parallel (compare the cross-hair images) 
have morphologies that are very different in the red and blue wing
data. The significance of this finding
is discussed in section~\pref{sec:phenomena}.

\subsection{Comparison of H$\alpha$ and \ion{Ca}{2} 854.2nm data }
\label{sec:hvsca}

Our time series dataset contains several instances where \ion{Ca}{2} images were
acquired within 10 seconds of H$\alpha$ images.  
The \ion{Ca}{2}
images were taken at $\pm0.06$ nm (equivalent to 
a Doppler shift of 21 \velu{}) compared
with $\pm0.11$ nm (50 \velu{}) for H$\alpha$.   
The interpretation of these data is generally complicated not only 
by the
non-LTE formation of the lines through the 
dependence on absorption strength on $n_i$ 
\citep{Athay1976},
but also by the fact that the observability of fibrils 
depends on both line widths and flow speeds (see our equation~\pref{eq:1}).
Nevertheless we can make a simple hypothesis
and test it using the parametric model of line formation of section
\pref{sec:model}.  

Some authors
\citep[e.g.][]{dePontieu+others2009,dePontieu+others2011} have
suggested that chromospheric plasma is accelerated and heated along
fibril tubes to some 50 -100 \velu{}.  
Let us assume that flows along
tubes occur (whether accelerated, decelerated or constant). How would
such 
flows be manifested in a comparison of H$\alpha$ and \ion{Ca}{2} data at
+0.11 and +0.06 nm?   
In the line cores there is much
confusion in the images making such a comparison difficult.  However,
in the cleaner wing images, even though they are acquired at
significantly different equivalent Doppler shifts
 (50 vs. 21 km/s),
we might expect to see
evidence of simply connected flows using the two lines, if they are present.   
The thermal properties of the two lines (excitation energies, 
line widths) are somewhat different. 
but chromospheric plasma is dense enough 
that momentum changing collisions between atomic species 
rapidly drives dissimilar bulk velocities to zero (making $s$ and
non-thermal contributions to $w$ similar for H and Ca).  
Thermal line widths and $n_i\Delta$ are very different for the lines, so
we certainly do not expect to see one-to-one correlations. 
But we should be able to find at least {\em some} clear
examples where a dark feature seen in the \ion{Ca}{2} line co-aligns
with a dark feature in H$\alpha$. 

To search for flow signatures in 
fibrils we co-registered (to sub-pixel accuracy) and subtracted 
\ion{Ca}{2} images from H$\alpha$ images, for 
three cases where data taken close in time (10 sec apart) are
available.  Results for two cases 
are shown in  Figure~\pref{fig:acceleration}. The two left
columns are speckle reconstructed images at +0.11 and +0.06 nm
respectively for H$\alpha$ and \ion{Ca}{2} 854.2 nm. The 
second row
contains data obtained 
some $\approx 200$ sec apart.  Two properties
are clear from these images: first, essentially no fibril structures are the
same between 14:12:03 UT and 14:15:26 UT- the lifetimes of the
observed 
fibrils is less than 200s.  Second, although some
features appear common to both H$\alpha$ and \ion{Ca}{2} data, most of
the time the H$\alpha$ and \ion{Ca}{2} fibrils are spatially separate 
from one another, even though at a lower angular
resolution ($\approx 1\arcsec$, say) 
one might conclude that arise from one and the same feature.   (Fibrils in the
lower of the two circles are examples).  To account for an offset of
say $0\farcs5$ by real lateral motion of fibrils 
would
require physical velocities $\gta 36$ \velu{}. Such displacements would be visible
in the difference image (rightmost column) as black and white features
aligned along the fibril lengths.  The lower circle includes one such
possible feature.  But most of the black/white adjacent features 
in fact correspond to differences
in brightness of faculae and remain present in both sets of data
separated by 200s (see the region near $x,y=836\arcsec,393\arcsec$).  

The difference images (rightmost panels in
Figure~\pref{fig:acceleration}) can also reveal 
plasma {\em accelerations}, which in the model are $d s/ d\ell$ which would be manifested as 
fibrils that show a moving absorption along their
lengths $\ell$.  
Our angular resolution $\sim 0\farcs2$ corresponds to
about 150 km on the Sun, so it is not unreasonable to expect to see
evidence of plasma acceleration along a typical fibril of length $5\arcsec{}
\equiv 3500$ km, if it is common.  
The thermal properties and the  particular acceleration profile 
will determine exactly where along a given flux tube the absorption will occur
in both lines. Therefore we expect to observe a variety of 
bright and dark features 
aligned along a tube in such images. If part of a simply accelerated
tube of plasma, their axes must be precisely
aligned in the very narrow, essentially unresolved fibrils.  
Remarkably, there are very few clear examples which might correspond to accelerated
plasma (Figure~\pref{fig:acceleration}). 
The upper circled region  shows an example of an {\em almost} co-aligned
transition from dark to light in the rightmost upper image.  
Close inspection reveals that the dark and light
regions  are 
not precisely co-axial, so those observed seem to appear by chance.
We have found no convincing case  in our
dataset, a factor of 70 less than the total number of fibrils
observed.

We estimate the likelihood that such alignments occur by chance.  Fibrils occur
in groups within which they tend to follow the
direction of their neighbors (refer to any IBIS image in the present
paper).  
We imagine dropping a ``Ca fibril'' of
length $5\arcsec$ (50 pixels) onto a given fibril clump seen in
H$\alpha$.  Assume that one H$\alpha$ fibril lies along a line 
between $x=0$ and
$x=5\arcsec$.  The base area of the fibril group appears to be on the
order of $A\sim
25$ square seconds of arc (Figure~\pref{fig:average}), from which
$\sim10$ fibrils originate.  The chance of finding one ``Ca
fibril'' with either end point within an area $a$ of 1 pixel by
1$\arcsec$ ($a=0\farcs1^2$), at either $x=0$ or $x=5$, is roughly $P_1
= 4a/A \sim 0.02$.  If fibrils have azimuthal distributions within a
group with a width of say 0.2 radians, only a fraction $P_2 = b/ 0.2$
will appear sufficiently aligned with the H$\alpha$ fibril.  For $b$
we use one half of the ratio of fibril width/length, $b\sim1/20$.  The
net probability for finding one coincidence {\em at random} is then
roughly $P=n\times n P_1 P_2 $ in one group which contains $n=10$ Ca
and $n=10$ H fibrils, or $P=0.4$.  With the $\sim 10$ groups in the
FOV (Figure~\pref{fig:average}) we expect to see at $\sim 4$
coincidences across the observed field of view at any given time. 
Given the roughness of
these estimates, we conclude that the lack of an observed relationship
between the \ion{Ca}{2} and H$\alpha$ fibrils, seen at 0.06 and then 0.11
nm respectively, is consistent with the null hypothesis. No fibrils
are detected above the number expected from a random alignment of 
\ion{Ca}{2} and H$\alpha$ 
fibrils.

Lastly, we made further comparisons of the H$\alpha$ and \ion{Ca}{2}
lines from the wavelength scans.   
The time series datasets analyzed below and by JRC12 
were obtained mostly at $\pm0.11$ nm in H$\alpha$, $\pm 0.06$ nm in \ion{Ca}{2} (see
Figure~\pref{fig:timeplot}).  From the wavelength scans we compared
first H$\alpha$ data at -0.103 nm to Ca II images taken at -0.136 nm,
i.e. at the same Doppler shift of $\approx 47$ \velu{}.  Each set of
H$\alpha$ and Ca II images were taken within 7 seconds of each other
and were co-aligned.  The fibrils observed in
the H$\alpha$ images are absent in \ion{Ca}{2}
images at this same Doppler shift.    It seems likely that
the optical depth  $\delta\tau_x$ in the \ion{Ca}{2} lines 
must be smaller than that of H$\alpha$ in the structures leading to 
the H$\alpha$  fibrils.  Thermal
broadening can in part explain this difference \citep{Cauzzi+others2009}, since for hydrogen
lines when $T_{ion} \sim 2\times 10^4$K, thermal broadening is on the
order of $w \approx 12$ \velu{}, 1/4 of the observed values of 47
\velu.   There may
also exist a contribution from the different thermodynamic states of
hydrogen and calcium in the fibril plasma that determine $n_i$ (see
e.g. \citealp{Athay1976}), but in section~\pref{sec:interp} we find
that variations in $n_i\Delta$ are important only for the \ion{Ca}{2}
line.

\subsection{Relationship of the fibrils to chromosphere, transition region and corona}

Figure \pref{fig:loops} shows the spatial relationship between the H$\alpha$ wing
fibrils and data of the low chromosphere (160 nm), transition region
(30.4 nm) and corona (19.3 nm) as seen with the AIA instrument on SDO.
AIA's angular resolution of 1\farcs5 is 7 times
lower
than the resolution of IBIS. Importantly, 1\farcs5 exceeds the widths
of many of IBIS  fibrils
by a factor of several.  Thus we can compare SDO and IBIS  images only at 
course angular resolutions. 
Earlier figures and Figure~\pref{fig:loops} shows that these wing fibrils originate from 
bright patches in the chromospheric network,  a well documented 
result \citep[e.g.][]{Athay1976}.  As such, they are also
grouped largely over the network where the 30.4 nm emission in
\ion{He}{2} is also brighter, as helium emission is associated with 
the chromospheric network \citep[e.g.]{Jordan1975}. 
While no 1:1 relationship with \ion{He}{2}
is expected (e.g., \citealp[][and references
therein]{Pietarila+Judge2004}), clearly both the fibrils
and 30.4 nm emission originate in  network regions.  In stark contrast,
the coronal emission is confined largely to active region loops
anchored in and around the sunspots.  Whatever the physical process is
that generates the  fibrils observed in the wings of H$\alpha$ 
and \ion{Ca}{2}, it has no detectable counterpart in
the corona
at this spatial resolution.
Any correlation with the \ion{He}{2} image at 30.4 nm,
although stronger, is still very weak: the lower left panel of
Figure~\pref{fig:loops} has fibril circles always with moderately
bright \ion{He}{2} emission, but there are many such bright regions
with no fibrils. 

If the H$\alpha$ fibrils are directly correlated with Type II spicules, 
it is difficult to 
reconcile the idea
that the fibrils observed
here have much to do with the supply of mass, momentum and energy into
the corona \citep[cf.][see the Discussion below]{dePontieu+others2009}.

\section{Discussion}

\subsection{New observational results}

In summary:
\begin{itemize}

\item{} At any time there are $\sim 120$ fibril-like
  structures seen at $\pm 0.11$nm either side of H$\alpha$ line center
  in the IBIS FOV. If covering the entire Sun there would be a total
  of $5\times10^4$ fibrils.
\item{} Of these, one in five exhibit the very rapid
  changes reported earlier (JRC12). 
 \item{}  The statistics of blue vs red wing data 
($\pm 0.11$nm) of H$\alpha$ are  different in
  this particular limb region.  
   The center of the IBIS frames has a line-of-sight that is at
  $\vartheta_C = 78^\circ$ to the local vertical on the Sun,
   so that horizontal flows yield higher fibril visibilities in the
   line wings than vertical flows of the same magnitude through
   Doppler shifts $s$ in equation~\pref{eq:1}. 
\item{}  The fibrils seen in red- and blue- 
wings in H$\alpha$  appear in
  the same regions of network with enhanced line widths, 
  but they are spatially and temporally
  distinct.  
\item{} Near-simultaneous  \ion{Ca}{2} and H$\alpha$ data obtained at
  different Doppler shifts (20 and 50 \velu{} respectively) reveal 
  no clear examples of acceleration along the
  length of a 
  fibril, of some 70 statistically independent cases. These statistics are
  consistent with chance alignment of fibrils seen in 
  \ion{Ca}{2} and H$\alpha$ data. 
\item{} The wavelength scans of H$\alpha$, although 
containing significantly more residual atmospheric seeing, nevertheless 
show that fibril profiles statistically are broader,
  shallower and more asymmetric than the mean   profile.
The red- and blue-wing  fibrils occur in areas of  broad H$\alpha$ and
Ca II line profiles, they are pervasive around magnetic concentrations.
\item{}  The red- and blue-wing fibrils seen in the 
  reconstructed images occur in areas that 
are correlated to bright UV/EUV emission from
the chromosphere and transition region seen at a lower resolution 
from space. We could find
no correlations with the brightness of the overlying corona.
\end{itemize}

\subsection{Interpretation of line wing images}
\label{sec:interp}

 The line profiles can be examined in the light of 
 the simple model of section~\pref{sec:model}.  
   Figures~\pref{fig:profiles} 
and \pref{fig:profilesc} show images of fibril profiles divided by the
average profile $r_{x,p}=I_{x,p}/\avq$.  The data shown, the first of their respective scans, 
are typical.  If we adopt
 $\avq$ as
the profile in the absence of overlying fibril absorption, then 
 with 
equation~(\pref{eq:1}), $r_{x,p} = 1+ \delta I_{x,p}/\avq $.  
The striking vertically banded structure seen in these Figures
is common to 
all fibrils, it stands in contrast to the two
non-fibril (``null'') cases also shown. 
A few 
fibril  profiles  are almost symmetric around
 line center.   In the model, these profiles 
correspond to terms proportional to 
 $\partial^2\phi_{x}/\partial x^2$ (see the fibril with index
17, for example), which therefore correspond to changes in the line widths
$w$.  

The majority of fibrils, however,  
show profiles proportional  to the first derivative as well. Some of these
profiles change along the fibril length as the two derivative components
compete for dominance (see those labeled 2 and 5 for example), 
indicative of acceleration or
deceleration ($ds/d\ell \ne 0$ as a function of distance $\ell$ along
the fibril length).  Thus we conclude that 
{\em most
fibrils, as measured using these seeing-influenced data, are broader
and Doppler shifted relative to 
the average profile}.  

To extract more information contained in these profile ratios
$r_{x,p}$, 
we have applied Principal Component
Analysis \citep[``PCA'', e.g.][]{Jolliffe2002} to the spectra.
PCA reveals the patterns present in the data and their 
relative frequencies of occurrence.  
For each pixel $p$ the vector
($r_{x,p}$) was constructed as a
function of frequency $x$, and the 
vectors over all pixels were averaged.  The outer product of this
average vector was
computed along with eigenvalues and eigenvectors of the derived 
matrix.  
Results of PCA of the H$\alpha$ fibril 
profiles are shown in Figure~\pref{fig:pcah}.  The eigenvalue
spectrum drops to less than 1\% after the first four components.
The same plot for the null pixels (not shown) is far steeper, dropping
to 0.01\% in the first four components.   The eigenvectors with
smaller eigenvalues have more noise, judging by the higher frequency
variations.  The data  are therefore represented adequately
(i.e. above noise levels) 
by a few principal components.  The figure shows 
the average profile $I_x$ and the eigenvectors
of the first six components of the ratio of $I_x$ to this average profile.  We recognize vectors 1 and 2 as second
order-like component ($\partial^2 \phi_x/\partial x^2$), and vectors 3 and 4 as 
first order-like components.  Higher order vectors 5 and 6 are 
associated with noise or systematic errors.   Vector 1 is a
combination of a DC offset and a second order-like profile, appearing
similar to the average $\delta I_x$ profile.  An offset (i.e., eigenvector
independent of wavelength index on the abscissae) reflects changes
independent of wavelength in the dataset- these are simple 
brightness variations of the photosphere underlying the fibrils.  

The PCA analysis therefore confirms that the 
H$\alpha$ fibril ratios $r_{x,p}$ therefore are dominated by a linear
combination of the zeroth, first and second derivatives of the line
profiles with wavelength.    According to
our model, the {\em
  H$\alpha$ fibril properties involve variations in all three parameters: brightness, 
line shift  and line width.   }  The null profiles are
similar but the eigenvalues of vectors 2,3,4\ldots{}  are all 0.1\% or
smaller- i.e. the occurrences of variations attributable to 
line widths and shifts are far smaller.  

Figure~\pref{fig:pcaca} shows the same plot but for \ion{Ca}{2}
fibrils. These have a similar eigenvalue
spectrum to H$\alpha$ but the first two eigenvectors appear similar to 
$\phi_x$, i.e. the brightness or 
zeroth derivative of the profile.  A mix of first and second order derivatives are in
components 3 and 4 with eigenvalues 0.05 and 0.03 respectively.   
Thus, {\em the \ion{Ca}{2} fibril variations correspond, in the model, 
first to changes in column density $n_i\Delta$,
second to changes in line width $w$ and third changes in line shift $s$ respectively. }

It is important to note that the lack of spatial correspondence
between the red and blue fibril images (section \pref{sec:rvb}) results primarily  from 
the asymmetry induced by changes in $s$. 

\subsection{Our work in context: Phenomenology}
\label{sec:phenomena}


The combination of high cadence, high spectral and angular resolution
of our chromospheric data approaches the limit of modern
instrumentation, and comparable datasets are still few and far between
(Table~1).
Early literature on spicules is important
but was based on observations with smaller ground-based telescopes in
the era before adaptive optics and image processing were available
\citep{Roberts1945,Beckers1968, Beckers1972}.  Such early studies
would not have detected the fine structures examined here
\citep[][see also our Appendix B]{Pereira+dePontieu+Carlsson2013}.  
To date, 
space-based data
\citep[e.g.][]{dePontieu+others2007} have been obtained with slower
cadences 
at single wavelengths and
in broader bandpasses than reported here.   Modern
ground-based work has usually emphasized image quality at the expense
of high cadence (Table 1).

\cite{Rutten2006,Rutten2007} reported very fine ``straws''
as long, bright and rapidly evolving features
seen near the limb in broadband \ion{Ca}{2} H images from the Dutch Open Telescope.  
Using data from {\it Hinode}, again in broadband \ion{Ca}{2} H images,
\cite{dePontieu+others2007} reported  two types -- at least -- of limb
spicules.
The ``type I'' spicules have been identified with the dynamic fibrils
seen on disk and explained as shock-driven plasma undergoing an
upward-downward motion cycle \citep{Hansteen+others2006}, while the
``type II'' spicules had many similarities with the ``straws'',
including a very large aspect ratio (widths of a few hundred km, lengths
up to several Mm), a short lifetime (10-150 s), and large amplitude
swaying presumably induced by pervasive Alfv\'en waves in the
chromosphere.
Several groups
have tried to identify these features on the solar disk, concluding
that they correspond to ``rapidly blue-shifted excursions'' (RBEs) in
H$\alpha$ and \ion{Ca}{2} 854.2 nm profiles, seen as blue-ward absorption components in
disk spectra
\citep[][]{Langangen+others2008,Rouppe+others2009,Sekse+others2012}. 
From their derived Doppler velocities and widths, these RBEs displayed a trend of acceleration and heating from their foot-points (in the magnetic network) upwards.

By early 2012 the combination of type II spicules, RBEs, and some EUV
observations from space
\citep{dePontieu+others2009,dePontieu+others2011} comprised an
apparently coherent picture of chromospheric plasma undergoing heating
and flowing into the corona in tube-like structures.
(However, in view of the large flux-tube expansion with height, it still
remained unclear how such plasma features could appear so collimated
so high above the solar surface, see JRC2012).
The story continues with the recent report of data
\citep{dePontieu+others2012} which apparently need
yet another
kind of motion within the spicules, namely a torsional motion, in
order to explain the observed slope of spectral lines at the limb.
Expanding on this idea, and re-analyzing some of the same datasets
used in earlier works, \cite{Sekse+others2013} highlighted ``rapid
redshifted excursions'' (RREs) in H$\alpha$ and \ion{Ca}{2} 854.2 nm
profiles seen close to disk center, with an occurrence about half that
of RBEs.  \cite{Sekse+others2013} claim that ``the striking similarity
of RREs to RBEs implies that RREs are a manifestation of the same
physical phenomenon''. As some RRE and RBE pairs appear adjacent to
one another, these authors interpret such neighboring signatures as
further evidence for torsional, vortex flows inside tubes.

To explain the behavior seen in the light of all these new 
datasets of fibrils, the traditional picture of plasma flows
in a tube apparently 
requires an increasingly complex array of physical processes
taking place within it, including outflow; heating; tube waves;
reflection of tube waves to produce standing waves
\citep{Okamoto+dePontieu2011}; vortex-like rotational motion
\citep{dePontieu+others2012,Sekse+others2013}.
Even so, it must be remarked that some of the RBEs/RREs
characteristics observed in recent high quality data sets (apart from
our own) are objectively difficult to accommodate into this picture:
Fig. 4 of \cite{Sekse+others2013} for example shows that a large
fraction of fibrils around a network patch displays larger Doppler
velocities and widths towards their footprints, quite contrary to what
required by the acceleration scenario. Even if swaying and/or
torsional motions are present within the fibrils, the large spatial
coherence of such flows (7-10,000 km in the horizontal direction,
i.e. along a string of network points) would require an organized
collective behavior wholly unexpected from first principles.  
Further, while the
presence of RRE and RBE pairs adjacent to one another can certainly
indicate torsional motions \citep[which were hypothesized in spicules
already by][]{Beckers1968}, it also implies that the horizontal
extension of these fibrils is quite larger than usually assumed when
looking at a single wavelength. The several examples displayed in
Figs. 2, 8 and 9 of \cite{Sekse+others2013} all suggest that the
fibrils involved have a width of 1 arcsec or even larger, which runs
contrary to the original definition of type II spicules
\citep{dePontieu+others2007}.

How do these pictures relate to our data?  
If we assume that torsional motions are
responsible for observed close 
and near-parallel pairs of red and blue wing
images (Figure~\pref{fig:rvb}),   then the data presented here 
show that 
conditions must be systematically different for the blue-ward and
red-ward components of this torsional motion.  
If separated by a distance of $\ell \sim 200$ km (see 
the parallel but offset red- and blue-wing data in the cross-hairs of 
Figure~\pref{fig:rvb}), the 
Alfv\'en crossing time between sides of the vortex would be a few
seconds (assuming a low $\beta$ plasma and $C^2_A \sim C^2_S
/\beta$).  This time is short compared with the lifetimes of most
fibrils examined, so
that a combination of MHD waves and rotational motion (with 
velocities $\lta C_A$) might be
expected to produce more symmetric pairs of fibrils than we 
observe, as the same material flows around the vortex becoming visible in 
both red and blue images.

Given this 
rapidly evolving literature, we can only speculate how our
data might really fit into it.  
On the one hand, the statistics, appearance, lifetime of the fibrils
presented in this paper suggest that we are observing RBEs and RREs
and/or type II spicules, close to the limb on the solar disk.  Yet,
several other characteristics do not fit with the pictures proposed 
in the work cited above. For example, we rarely observe
clear cases of side by side blue/red fibrils (which could evidence
vortex-like motions), and for the few cases in which we do, we also
observe a significant spatial asymmetry in the blue/red images along
the fibrils' length, something  not reported by \cite{Sekse+others2013}.
Further, we do not find a
clear relationship between \ion{Ca}{2} 854.2 nm and H$\alpha$ fibrils
in the sense highlighted by \cite{Sekse+others2012}, i.e. that ``Ca II
8542 RBEs are connected to H$\alpha$ RBEs and are located closer to
the network regions with the H$\alpha$ RBEs being a continuation of
the Ca II 8542 RBEs''.  Certainly our data, obtained close to the
solar limb, favor detection of horizontal motions that can induce line
shifts, but this observational bias should not be of any relevance for
the detection of torsional motions within a (mostly) vertical tube, or
of accelerated plasma progressively appearing at larger heights in
projection, if present (Sect. \ref{sec:hvsca}).

We have no explanation for these discrepancies, and
just note that the interpretation  of data revealing 
chromospheric ``fibrils''
remains rather confused.
We suggest that the analysis of {\em full} spectral
profiles at the highest cadence and spatial resolution should be the
highest priority of future research: as we showed in
Sect. \ref{sec:scans}, the fibrils' profiles are most often broadened
with respect to the ``quiet'' areas around them, a result consistent
with older literature \citep[e.g.,][]{Athay1976} and with
the findings of \cite{Cauzzi+others2009}. Intensity changes at any given
wavelength (or few wing wavelengths) will depend on the original
profiles (Sect. \ref{sec:model}) so that interpretation only in terms
on separate red and blue ``components''
\citep[e.g.][]{Rouppe+others2009,Sekse+others2013} might be subject to
bias. In our spectral scans we find from PCA analysis of line profiles
that the origins of variations in H$\alpha$ that lead to wing fibrils
(and hence RBEs, RREs etc) are, to leading order, dependent on the
Doppler shifts and widths. 

\subsection{Conjectures and Refutations}

How can we then proceed, given the bewildering array of observational facts summarized above? 
In this paper we seek to {\em disprove} hypotheses - the approach advocated by
\cite[][``Conjectures and refutations'' is a famous monograph by Popper]{Popper1972}.  
Most  work on chromospheric fine
structure has not explicitly sought to do this. 
While far more mundane, than, say, the persistent precession
of Mercury's orbit, highlighted by Le Verrier and 
which refused to agree with Newtonian theory,
JRC12 presented data 
whose essential properties required some 
improbable mental gymnastics to make 
them fit into the tube model.
We
therefore appealed to sheets as an alternative
 physically plausible 
hypothesis needed to explain
critical aspects of the data, to be tested further.

The reader might justifiably adopt ``Occam's Razor'' and accept the ``simplest
answer compatible with the data''.    But there is a problem: what actually {\em is} the simplest answer
for chromospheric fine structure, given, for example, that there is no explanation even
for the extraordinarily large aspect ratios found? Let us consider the
two cases physically, putting
aside briefly the data presented by JRC12:

\begin{itemize}

\item{} The tube picture is a natural choice because the observed
  phenomena certainly {\em appear} like thin straws, and the
  electrically conducting plasma (partially or fully ionized) is
  frozen to tubes of magnetic flux on observable scales.  But magnetic
  structure rooted at the network photosphere boundary appears mostly
  like rivers not isolated tubes
  (JTL11- see the references therein). 
  Also, there is as yet
  no straightforward explanation for the formation of such thin
  structures (aspect ratios $>10-20$) that do not measurably expand with
  height within the chromosphere\footnote{The much ``fatter'' dynamic
  fibrils associated with spicules of ``type I'' are a quite different phenomenon 
  with much smaller aspect ratios; see \citealp{Hansteen+others2006}
  and appendix C.}.

\item{} The sheet picture was prompted through consideration of ribbon-like 
  observed photospheric network magnetic structure, 
  of interfaces between bundles of flux
  \citep{vanBallegooijen+others1998,Priest+others2002} and of Parker's
  theory of the formation of TDs.  In the latter, thin sheet-like structures are
  predicted to originate simply because force balance and magnetic
  topology are not generally consistent with continuous solutions for
  the magnetic field \citep{Parker1994}.  They are weak solutions to
  the governing partial differential equations similar to shocks so familiar in continuum fluid
  mechanics.  The magnetic field direction, but not magnitude, changes
  slightly across flux surfaces.  The apparent width of plasma embedded in
  sheets can readily be below
  observable scales (appendix B), so that the visible warps in sheets do not need to expand with
  height as tubes of potential fields must do.

\end{itemize}

\subsection{Data vs models and some physical implications}
\label{sec:physics}.

The idea that spicules supply mass, momentum, and energy to the corona
is at least 65 years old
\citep{Thomas1948a,Miyamoto1949a,Pneuman+Kopp1978,Athay+Holzer1982}.
Naturally, spicules of type II, which share qualitative similarities
of upward apparent motion and fading
\citep{Pereira+dePontieu+Carlsson2013} with the ``classical'' spicules
reviewed by \cite{Beckers1968}, continue to draw attention as a
manifestation of mass and energy supply to the corona.  Much recent
work has used spacecraft data from chromosphere to corona to 
support this picture 
\citep{dePontieu+others2007,
  dePontieu+others2009,dePontieu+others2011}.  As such, essentially
all workers interpret the data found in terms of the motion (flows 
and wave motion) in elemental {\em tubes} of plasma frozen to magnetic
lines of force.  Do our data allow us to try to falsify this or the
sheet-like proposition discussed in the Introduction?

If, as appears reasonable, we are observing the near-limb
counterparts  of RREs, RBEs and type II spicules, the lack of
spatial (let alone temporal) correlation between the fibrils and the 
overlying coronal emission is striking and important. 
Either the particular features we are observing are unrelated to
those that have been reported to accelerate heated plasma into the corona 
\citep[e.g.][]{dePontieu+others2009,dePontieu+others2011}, being 
a separate, mostly horizontal population of fibrils/spicules, or the 
acceleration proposal needs further
examination\footnote{Interestingly, the  Doppler widths are not
incompatible with the picture  of \cite{Judge+Carlsson2010}, who
required supersonic motions to explain the curious 
absence of absorption across the limb in 
\ion{Ca}{2} images.}. 

It is critical to the ``tube'' hypothesis that standing waves account
for the very fast phase speeds we have reported.   
Can we reject the hypothesis that 
the standing\--wave scenario explains the data we have
reported here and earlier, without reference to the many other
studies? 
In JRC12 we dismissed standing waves 
for the particular behavior found in that paper because complete fibrils
appeared out of nowhere, with no evidence of the setting 
up of a standing wave from the interaction of two propagating waves.
In appendix A we review these and other  problems with standing waves.  
But the standing wave scenario can, in fact,  indeed be rejected for
our data.
One
primary characteristic of the time series measured in one very narrow
spectral band is that
the absorbing intensities are remarkably uniform along the entire length of the
fibrils (see the figures in JRC12).  
If wave motions were responsible, the kink/helical  
waveforms would have Doppler shifts that depend on position along the
tube, having nodes and antinodes somewhere along their length.
This must translate, for a harmonic wave, to absorption features
which track in space loci where Doppler shifts match the 
observed bandpass- we should observe evidence of 
this nodal structure along the fibril
length.  The calculations of section \pref{sec:model} show just how
sensitive the intensities can be to small, subsonic changes in line
shift and width parameters. 
But 
we do not observe the gradients of intensity in these fibrils that
would correspond to the necessary varying velocities of plasma
organized in standing waves. While tubular in appearance, 
the fact that we fail to observe a critical component of the flux tube
model- the standing waves of plasma that supposedly travel along these
flux tubes- discredits the picture. 

Can we then reject sheets?  For spicules of ``type I'' there is no
debate (and never has been).  We know of no data that contradict the
model predictions of \cite{Hansteen+others2006}, so a new hypothesis
is not warranted there.  
However,
unlike \cite{Pereira+dePontieu+Carlsson2012},
we believe that evidence could and can be found against the sheet hypothesis
with current instrumentation (see Appendix C).  
Two critical tests would serve to reject sheets.

First is the
observation that ``type II spicules'' appear only to travel upwards.
Time reversal applied to a harmonically varying sheet such as shown in
JTL2011 means that as many fibrils (if sheets in projection)
will move down as well as up.  The type II spicule cannot therefore
be such a simple sheet.  In JTL2011 we noted that sheets will tend to
be driven by forces below them.  
Perturbations from below propagate up
in the form of Alfv\'enic disturbances.  In order to break the time
reversal symmetry there must be some hysteresis in these motions.
Alfv\'en waves are naturally damped by ion-neutral collisions, an
irreversible process that breaks this symmetry.  Such effects may also
account for some heating of chromospheric plasma.  It remains to be
seen the magnitude of this effect, it may prove sufficient to reject
the sheet hypothesis in some fibrils, in the absence of the needed
stereoscopic data.   

Second, if  imaging spectroscopy proves that 
vortex motions in these magnetically dominated plasmas 
are the rule, as reflected in the observation of
parallel 
pairs of blue- and red- shifted fibrils \citep{Sekse+others2013}
Occam's Razor would suggest that some rotational symmetry is 
required.  This is not expected to be a common form for tangential
discontinuities formed as anticipated in the spontaneous or tectonic 
pictures.  While ``solar tornadoes'' have been found
stereoscopically in much larger structures in the corona \citep{Nistico+others2009},
fibril observations so far have
shown that adjacent, parallel fibrils have Doppler shifts that might 
be vortices.  However, as
we noted (section \pref{sec:phenomena}), the red and blue motions are spatially 
asymmetric, and,
by enlarging the foot-point area  we see
that they have a good chance of being adjacent, unrelated flows (see the
calculation of chance occurrences of seeing continuous jets 
in Section \pref{sec:hvsca}). In the absence of shocks and
embedded tangential discontinuities, vortices demand
a {\em steady} gradient in speed as a function of distance
from the vortex axis.  Detection of the full profile of vortices for a
given observation would
serve to reject the sheet picture for the underlying structure.   This has not been
demonstrated.

%

What is the physical meaning of  the 
relationship between the fibrils we observed and the 
emission from the chromosphere, transition region and corona shown in 
Figure~\pref{fig:loops}?  If, as argued by several authors
\cite[e.g.][]{dePontieu+others2009}, there is
a relationship between type II spicules and the supply of mass and
energy to the overlying corona, we should see at least some correlation
of the occurrence of the fibrils with the overlying coronal emission.
No such correlation exists in our data, 
even though our data span over 30 minutes
which is roughly the cooling time of the overlying corona.  Also,
we are not
convinced by the analysis of \cite{dePontieu+others2011} who correlate 
various fibrils and spicules of type II with 
lower angular resolution, broad\--band SDO data.  In particular the
connections to the corona are very unclear, and \ion{He}{2} 30.4 nm 
emission is not understood \citep{Pietarila+Judge2004, Judge+Pietarila2004}. 
In our data 
the fibrils are correlated with chromospheric network including 
emission in \ion{He}{2}.  But there is much \ion{He}{2} emission
outside of the areas where we observe many fibrils (circled regions in
the figures).  
Such a correlation, even if it did exist, does not require volumetric 
heating of
the form implied by de Pontieu et al., since alternative explanations 
involving cross-field particle and heat transport from the corona to
the cooler spicular material have 
not yet been rejected \citep{Athay1990,Judge2008}.  In fact, the smooth
tailing off of intensity in the type II spicules with height might be 
accounted for naturally in such scenarios.  Even more surprising, if
the mechanisms discussed by \cite{Athay1990,Judge2008} are important,
energy is transferred {\em from} the corona {\em to} the chromospheric
plasma- precisely the opposite of the proposition of \cite{dePontieu+others2009}!

Lastly, at the risk of
precipitating a non-productive discussion, we feel we must comment on
some relevant criticisms of our recent work. We do this in appendices A,B and C.  The
first Appendix casts a critical eye on the proposal that standing waves are a
credible explanation for the fast phase speeds reported by JRC12.
The second discusses physical lower limits to the
scales across spicules in the light of work by
\cite{Pereira+dePontieu+Carlsson2013}.  The third comments on a
subsection of a paper entitled ``are spicules sheets''
\citep{Pereira+dePontieu+Carlsson2013}.  It is hoped that the
debate
can and will end by the rejection of a model when it
fails to reproduce critical data.  The sudden
appearance along the entire length of fibrils 
reported in JRC12, augmented with the present analysis, suffices to
reject tube waves as a plausible explanation for the 
significant fraction (one in five) of fibrils evident in our
data.

\subsection{Implications for future chromospheric observations}

Our results extend the broad (but, to us at least, confusing)
literature which reports very rapid evolution of chromospheric
structures consisting of narrow elongated structures variously called
fibrils, mottles, spicules, spicules of type II, RBEs, now RREs, jets, even ``cool
loops'' \citep[e.g.][]{Athay1976,
  dePontieu+others2007,Rouppe+others2009,deWijn2012,
  Judge+Centeno2008,Sasso+others2012,Sekse+others2013}.
  To understand the real
origin and importance of these structures it seems profitable to
pursue the highest possible cadence imaging spectroscopy of the Sun.
IBIS is one of several ground-based Fabry-Perot instruments that can
observe large, circular areas of the Sun ($\sim 10^4$ square
arcseconds) with useful reconstructed data at 1Hz, centered at a
single wavelength and a very high ($R=250,000)$ spectral resolution.
This stands in contrast to existing space-based instrumentation. IRIS
is the latest generation of UV ``imaging spectrographs'' in which a
slit is rastered rapidly to generate images.  The two techniques are
complementary.  Given our data, we choose  5 seconds as an upper limit to the
cadence for obtaining (almost) seeing-free spectral images. We also need
a field of view spanning roughly $10\arcsec\times10\arcsec$ to see complete
fibrils which are oriented independently of any 
instrumental detector. How do these two instrument configurations 
match up to these constraints? Consider:

\begin{itemize}
\item{} IBIS could
obtain five wavelengths across a spectral line in bursts of ten 0.1s
exposures, and obtain 100 such areas.  If polarimetry were needed,
the cadence would decrease to 20 seconds for the same conditions. 

\item{} In the same time IRIS could acquire far richer set of UV
spectral data, but the area covered is limited to $n\sim 10^2$ square
arc seconds where $n$ is the raster step size.  Since $n \sim 10$ is
needed to cover an area of width $3\arcsec$ with $0\farcs3$ steps, it
is clear that $n\sim30$ is needed.  Each step will typically require
1s.  Slit-based instruments actually are not ideally suited to the study
of the thin, dynamic fibril structures which have lengths of 5 seconds
of arc and which are fractions of an arc second across.  
\end{itemize}

Thus, a combination of imaging and slit spectroscopy is required. 
Both are available now. Both should be used to challenge  proposed
physical pictures of fibrils.

\acknowledgments IL is grateful to the REU program at MSU for support
of a visit to Montana State University during the summer of 2013.  PGJ
is grateful to the Directors of INAF, HAO and the Physics Department
of Montana State University for support of a visit to Arcetri 
and of a sabbatical
leave at MSU, during which this work was performed.


\figone
\figtwo
\figthree
\figmvs
\figprofiles
\figprofilesc
\figw
\figacc
\figloops
\figrvb
\figpcah
\figpcaca
\clearpage

\tabone

\clearpage
\appendix
\section{ Standing waves and spicules}

\cite{Sekse+Rouppe+dePontieu2013} 
cite 
\cite{Okamoto+dePontieu2011} (henceforth ``OdP2011'')
 as providing 
``conclusive'' evidence 
that standing waves are an
acceptable explanation of rapid changes seen along the
lengths of spicules.  Let us critically examine this claim, first as
applied to their data, and then in reference to our data.  

OdP2011 analyzed a time series of limb data of a coronal hole boundary
in the \ion{Ca}{2} H line with the 0.3 nm-wide filter of the SOT
instrument on the Hinode satellite.  They average 1.6s exposures
over 9 frames for a cadence of 14s.  Unspecified radial density
gradient and spatial high pass filters were applied.  Their algorithm
finds bright features in the time-averaged time series that are
persistent in time ($>45$s lifetime) and space (30 pixel overlap of
the feature's long axis between frames), long ($> 8\arcsec$ at its
maximum), have weak curvature ($< 5^\circ$ over lengths of
4$\arcsec$).  They identified 89 ``spicules'' meeting these and other
criteria, the bulk of them expected to be of ``type II''
\citep{dePontieu+others2007}\footnote{A type II spicule is at present
  defined only phenomenologically. The type I spicule has an
  acceptable physical interpretation \citep{Hansteen+others2006} in
  terms of slow shocks propagating along magnetic flux tubes.}.

The spicules so measured are projections of the total line intensity,
integrated over the SOT bandpass and along the line of sight, with
intensities modified by an unspecified radial function.  We assume the
brightness drops by typical values (a factor of 10 in about 8$\arcsec$,
\citealp{Bjolseth2008}).  Here are the basic assumptions made by 
OdP2011:

\begin{itemize}
\item The unperturbed spicule consists of a  cylinder of cool
  plasma embedded in the hot corona. 
\item The brightness (and hence visibility) of the spicule is
  proportional to the density of the plasma to some positive  power
(they assume +1), with no dependence on other thermal conditions
 such as plasma temperature and velocity. 
\item The brightness does not depend on multiple photon scatterings - it is optically thin.
\item The observed motions are displacements perpendicular to the
  spicule axis.
\item Such motions must therefore be 
  tube waves of Alfv\'enic character (magnetic tension is the restoring
  force). Kink modes are the most readily identified by their algorithm. 
\item  Phase speeds exceeding tube speeds are attributed to wave
  reflection from the transition region  above the spicule ``top''.
\end{itemize}
At long wavelengths the propagation speed is the 
kink speed $C_K$ \citep[eq.~8 of ][]{Nakariakov+Verwichte2005} which is
the density weighted average of the Alfv\'en speeds internal and
external to the tube:
\begin{equation}
C_K^2 = \frac{B_0^2 + B_e^2 }{ 4\pi (\rho_0+\rho_e)}
   \end{equation}
The internal density $\rho_0$ exceeds the external density $
\rho_e$ if spicules are high density, low temperature intrusions
into coronal plasma.  With $B_0 \lta B_e$ (plasma pressure inside
the tube exceeding those outside) $C_K \lta 2 B_0^2/4\pi\rho_0$.
By assuming $\rho_0 \propto $ brightness$^\alpha $ where
$\alpha=1$, OdP2011 conclude that this explains the observed
increase in propagation speed along the spicule length.
 
This picture has the following problems:  (1) The scale of variation
of the propagation speed are of the same magnitude as the spicule
length and wavelength of the oscillations.  Oscillations will interact
with these gradients in the ``background state'' in a complex fashion
(non-WKB continuous reflection, mode conversion), not merely propagate
unhindered to the spicule ``top''.  (2) Second, if high phase speeds
result from reflection, one must have oppositely directed propagating
waves of similar amplitude that are phase coherent with one another
for at least one cycle.  Two such waves would have to occur by chance,
because even if 100\%  of the wave energy is reflected at 
a transition region, the non-WKB effects cause amplitude and phase
changes along the spicule length.  (3) Third, a 
steep brightness gradient (scale lengths 100 km or less) 
should be observed in at least some spicules, yet almost all show a
slow fading of brightness. 

These points must be addressed if we are to believe the interpretation
put forward by OdP2011, and the very strong statement of 
\cite{Sekse+Rouppe+dePontieu2013}. 

Together with the objections concerning the Doppler signatures of
standing waves 
raised in the Discussion, 
we must conclude that OdP2011 have not ``conclusively shown'' that standing
tube waves are the appropriate physical picture for their
observations. Further, as we have reported earlier, our observations have features that
tube waves cannot describe.

\section{The smallest scales of spicules}

In a long overdue paper, \cite{Pereira+dePontieu+Carlsson2013} tie the
sub-arcsecond data to the earlier literature on spicules by degrading
their 0.32 nm passband \ion{Ca}{2} data from Hinode.   
\begin{quote}
  ``These results illustrate how the combination of spicule
  superposition, low spatial resolution and cadence affect the
  measured properties of spicules, and that previous measurements can
  be misleading.''
\end{quote}
Unfortunately the same argument can be applied to {\em all}
observations, including theirs.  
To proceed, we must ask if there are  {\em physical} reasons to
expect that modern telescopes are resolving anything on fundamental
scales.  Considering the chromosphere as a partially ionized
magneto-fluid \citep[e.g.][]{Braginskii1965}, the smallest physical scales 
are set by transport coefficients, assuming that the driving motions
can lead to small scales, as argued by \cite{Parker1994}. The electrical conductivity and
viscosity tensors are the most relevant coefficients.  In all spicule
models the spicule structure  is maintained by the magnetic field
\citep{Sterling2000}.   The time
$t$ taken for plasma to diffuse across a scale $\ell$ is simply
\begin{equation}
 t \sim \ell^2 / \eta
\end{equation}
where $\eta$ is the relevant component of the diffusivity  tensor.  We
select the Pedersen conductivity since physical gradients scales across the
field in spicules are presumably far larger  than along them.  If 
we adopt $t \sim
100$ s, the thermodynamic lifetime of spicules, as a minimum for the
magnetic lifetime, then features larger than $\ell \gta \sqrt{100 \eta}$
will maintain their integrity (magnetic field frozen to plasma).  
%
%
%
%
%
%
With kinetic values of $\sigma \sim 3\times 10^{10}$ s$^{-1}$ for
plasma near the top of the chromosphere \citep{Goodman2004}, we find
$\ell > 5$ km.  Below this scale we would expect the frozen field
condition to fail, meaning that a spicule would lose its thermal
identity within its lifetime.  Turbulent conductivity would have to be
some 1000 times less in order for $\ell $ to approach observable scales near 150
km.  Using the estimate of ``Bohm'' diffusion ($\sim \omega\tau$ times
the kinetic value, given by Braginskii 1965, where $\omega$ is the
proton gyro frequency and $\tau$ the mean proton collision time with
all other particles), this would require $\omega \tau \sim 1000$.
\cite{Goodman2004} finds values closer to 10 in the upper
chromosphere.

We conclude that the very arguments presented by 
\cite{Pereira+dePontieu+Carlsson2013} should be expected to apply to
their own 
among all other current
data.  They present no credible reasons why we should expect
Hinode (or any other instrument in existence) to have really resolved
solar spicules. We urge caution before making definitive statements about chromospheric dynamics.

\section{Comments on: Are spicules sheets? }

Here we reject scathing criticisms of our work in section 5.5,
``are spicules sheets'' of yet another article on Hinode \ion{Ca}{2}
``type II spicules'' by \cite{Pereira+dePontieu+Carlsson2012}.   We
have presented evidence that the behavior of some structures is indeed
{\em incompatible with the flux tube picture} (JRC12, this paper).
Thus we are convinced that the tube models cannot, without mental
gymnastics and implausible physical assumptions that we have
presented, explain these particular data.  This is {\em sufficient} to
warrant what the authors call ``a paradigm shift regarding the
spicule deposition of energy and mass in the corona'', indeed,
niggling discrepancies and creative maneuvers to sidestep them are
often signs that something is fundamentally incorrect.  This is how science
progresses \citep{Kuhn1970,Popper1972}.

In the following we quote directly from their paper and respond to
each point in turn.
\begin{quotation}
  ``\ldots in general we fail to see any significant observational
  proof of spicules as sheets''
\end{quotation}
Seeking observational proof is philosophically questionable at best
\citep{Popper1972}, and naive in the complex case of remotely sensed
data for the solar atmosphere.  The nearest thing to a proof will be
the stereoscopic observation of the chromospheric fine structure
which will require new and
challenging observations from a very different vantage point of the
earth (JTL11).  Turning it around, they continue
\begin{quotation}
``\ldots  in the vague definition formulated by JTL11, it
will be very difficult to find conclusive observational evidence for
or against this hypothesis, because of the unknown and invisible
orientation of the sheets\ldots''
\end{quotation}
In JTL11 it is clearly specified that 
the sheets are supposed to form in {\em tangential discontinuities} in
which there is a dominant guide field (as in a tube) but in which the
field direction changes slightly as one passes through the sheet.  
The guide field is entirely equivalent to the unperturbed tube field
in the tube picture.  These fields have already been measured in some
spicules \citep{Lopez-Ariste+Casini2005,Centeno+others2010}. 
It
is precisely the reconnection of the tangentially discontinuous field
that  drives motion perpendicular to the guide field within
the sheet.  This motion bodily moves plasma across the guide field
direction \citep[Figure 2 of][]{Judge+Tritschler+Low2011}.  Coupled
with 
eventual heating as the kinetic energy is dissipated by viscosity,
this naturally explains why chromospheric material can appear at
coronal heights.   
We know that very few fibrils are vertical, most have a ``guide
field'' that has a horizontal component.  Reconnection of the
tangential 
component will lead to observable Doppler shifts corresponding to 
both horizontal and vertical motions.   
If such motion occurs upwards out of the chromosphere, it can proceed
to significant heights almost unimpeded. If downwards, the
motion faces a steep uphill battle to continue as it ploughs into the
stratified chromosphere with a density scale height $z_\rho$ of only $10^2$
km. Such a gradient will serve to turn downward motions into upward
motions within a time of at most $ z_\rho/C_S \approx 10$ seconds, with 
$C_S \sim10$ \velu{} being the sound speed. 
 
The idea that our suggestion
involves supernatural phenomena:
\begin{quotation}
``One issue that is not addressed at all in the sheet 
hypothesis is how the cool plasma contained in spicules ``magically''
appears at 
what are essentially coronal heights.''
\end{quotation}
denies one of the basic conundrums of chromospheric physics, since,  as stated by JTL2011, this is a
general problem for {\em any} spicule model:
\begin{quotation}
``Still unanswered is the important question: how does 
the Sun make such long cool structures with lengths say 40× the
hydrostatic pressure scale height (e.g., Sterling 2000)?''
\end{quotation}
Their dismissal of one of our motivations:
\begin{quotation}
  ``\ldots if type I spicules are not sheets and indeed jets, this
  throws away one of the main arguments for the very existence of
  sheets, which JTL11 summarize as “how can one get straw-like
  structures out of the fluted sheet fields that appear to dominate
  the photospheric network (\ldots)?” ''
\end{quotation}
is misplaced, based upon the very data and model for type I spicules
which the authors have presented earlier.   Type I spicules have a much
smaller aspect ratio (length/width $\sim$ 1-5, see e.g. figures in
\citealp{dePontieu+others2007b}), they are the limb equivalent of
dynamic fibrils \citep{Hansteen+others2006} in which shock driven
blobs of plasma ejected upwards along magnetic fields merely to return
to the surface in response to photospheric motions.  Type I spicule lifetimes
are also longer than those of type II. 
The aspect ratio
of our fibrils, the type II spicules, RREs and RBEs etc. is far
higher, 10-20 at least.  The type I spicules are not straw-like in any
sense- they are merely plasma that responds to compressional forcing
beneath like a fountain or geyser.  The type II spicules present a
very different set of problems, as stressed originally by 
\cite{dePontieu+others2007}.


\begin{thebibliography}{}

\bibitem[\protect\astroncite{Athay}{1976}]{Athay1976}
Athay, R.~G.: 1976,
\newblock {\em The Solar Chromosphere and Corona: Quiet Sun\/},
\newblock Reidel: Dordrecht

\bibitem[\protect\astroncite{Athay}{1990}]{Athay1990}
Athay, R.~G.: 1990,
\newblock {\em Astrophys.\ J.\/} {\bf 362}, 364

\bibitem[\protect\astroncite{Athay and Holzer}{1982}]{Athay+Holzer1982}
Athay, R.~G. and Holzer, T.: 1982,
\newblock {\em Astrophys.\ J.\/} {\bf 255}, 743

\bibitem[\protect\astroncite{Beckers}{1968}]{Beckers1968}
Beckers, J.~M.: 1968,
\newblock {\em Solar Phys.\/} {\bf 3}, 367

\bibitem[\protect\astroncite{Beckers}{1972}]{Beckers1972}
Beckers, J.~M.: 1972,
\newblock {\em Ann.\ Rev.\ Astron.\ Astrophys.\/} {\bf 10}, 73

\bibitem[\protect\astroncite{Bj\o{}lseth}{2008}]{Bjolseth2008}
Bj\o{}lseth, S.: 2008,
\newblock {\em Master's thesis\/}, Oslo University

\bibitem[\protect\astroncite{Braginskii}{1965}]{Braginskii1965}
Braginskii, S.~I.: 1965,
\newblock {\em Reviews of Plasma Physics.\/} {\bf 1}, 205

\bibitem[\protect\astroncite{{Cauzzi} {\em et~al.}}{2009}]{Cauzzi+others2009}
{Cauzzi}, G., {Reardon}, K., {Rutten}, R.~J., {Tritschler}, A., and
  {Uitenbroek}, H.: 2009,
\newblock {\em Astron.\ Astrophys.\/} {\bf 503}, 577

\bibitem[\protect\astroncite{{Cavallini}}{2006}]{Cavallini2006}
{Cavallini}, F.: 2006,
\newblock {\em Solar Phys.\/} {\bf 236}, 415

\bibitem[\protect\astroncite{{Centeno} {\em et~al.}}{2010}]{Centeno+others2010}
{Centeno}, R., {Trujillo Bueno}, J., and {Asensio Ramos}, A.: 2010,
\newblock {\em \apj\/} {\bf 708}, 1579

\bibitem[\protect\astroncite{{De Pontieu} {\em
  et~al.}}{2012}]{dePontieu+others2012}
{De Pontieu}, B., {Carlsson}, M., {Rouppe van der Voort}, L.~H.~M., {Rutten},
  R.~J., {Hansteen}, V.~H., and {Watanabe}, H.: 2012,
\newblock {\em Astrophys.\ J.\ Lett.\/} {\bf 752}, L12

\bibitem[\protect\astroncite{{de Pontieu} {\em
  et~al.}}{2007}]{dePontieu+others2007}
{de Pontieu}, B., {McIntosh}, S., {Hansteen}, V.~H., {Carlsson}, M.,
  {Schrijver}, C.~J., {Tarbell}, T.~D., {Title}, A.~M., {Shine}, R.~A.,
  {Suematsu}, Y., {Tsuneta}, S., {Katsukawa}, Y., {Ichimoto}, K., {Shimizu},
  T., and {Nagata}, S.: 2007,
\newblock {\em Publ.\ Astron.\ Soc.\ Japan\/} {\bf 59}, 655

\bibitem[\protect\astroncite{{De Pontieu} {\em
  et~al.}}{2011}]{dePontieu+others2011}
{De Pontieu}, B., {McIntosh}, S.~W., {Carlsson}, M., {Hansteen}, V.~H.,
  {Tarbell}, T.~D., {Boerner}, P., {Martinez-Sykora}, J., {Schrijver}, C.~J.,
  and {Title}, A.~M.: 2011,
\newblock {\em Science\/} {\bf 331}, 55

\bibitem[\protect\astroncite{{De Pontieu} {\em
  et~al.}}{2007}]{dePontieu+others2007b}
{De Pontieu}, B., {McIntosh}, S.~W., {Carlsson}, M., {Hansteen}, V.~H.,
  {Tarbell}, T.~D., {Schrijver}, C.~J., {Title}, A.~M., {Shine}, R.~A.,
  {Tsuneta}, S., {Katsukawa}, Y., {Ichimoto}, K., {Suematsu}, Y., {Shimizu},
  T., and {Nagata}, S.: 2007,
\newblock {\em Science\/} {\bf 318}, 1574

\bibitem[\protect\astroncite{{De Pontieu} {\em
  et~al.}}{2009}]{dePontieu+others2009}
{De Pontieu}, B., {McIntosh}, S.~W., {Hansteen}, V.~H., and {Schrijver}, C.~J.:
  2009,
\newblock {\em Astrophys.\ J.\ Lett.\/} {\bf 701}, L1

\bibitem[\protect\astroncite{{de Wijn}}{2012}]{deWijn2012}
{de Wijn}, A.~G.: 2012,
\newblock {\em Astrophys.\ J.\/} {\bf 757}, L17

\bibitem[\protect\astroncite{{Goodman}}{2004}]{Goodman2004}
{Goodman}, M.~L.: 2004,
\newblock {\em Astron.\ Astrophys.\/} {\bf 416}, 1159

\bibitem[\protect\astroncite{{Hansteen} {\em
  et~al.}}{2006}]{Hansteen+others2006}
{Hansteen}, V.~H., {De Pontieu}, B., {Rouppe van der Voort}, L., {van Noort},
  M., and {Carlsson}, M.: 2006,
\newblock {\em Astrophys.\ J.\ Lett.\/} {\bf 647}, L73

\bibitem[\protect\astroncite{Jolliffe}{2002}]{Jolliffe2002}
Jolliffe, I.~T.: 2002,
\newblock {\em {Principal Component Analysis}\/},
\newblock Springer, second edition

\bibitem[\protect\astroncite{Jordan}{1975}]{Jordan1975}
Jordan, C.: 1975,
\newblock {\em Mon.\ Not.\ R.\ Astron.\ Soc.\/} {\bf 170}, 429

\bibitem[\protect\astroncite{Judge}{2008}]{Judge2008}
Judge, P.~G.: 2008,
\newblock {\em Astrophys.\ J.\ Lett.\/} {\bf 683}, L87

\bibitem[\protect\astroncite{{Judge} and {Carlsson}}{2010}]{Judge+Carlsson2010}
{Judge}, P.~G. and {Carlsson}, M.: 2010,
\newblock {\em Astrophys.\ J.\/} {\bf 719}, 469

\bibitem[\protect\astroncite{Judge and Centeno}{2008}]{Judge+Centeno2008}
Judge, P.~G. and Centeno, R.: 2008,
\newblock {\em Astrophys.\ J.\/} {\bf 687}, 1388

\bibitem[\protect\astroncite{Judge and Pietarila}{2004}]{Judge+Pietarila2004}
Judge, P.~G. and Pietarila, A.: 2004,
\newblock {\em Astrophys.\ J.\/} {\bf 606}, 1258

\bibitem[\protect\astroncite{{Judge} {\em
  et~al.}}{2012}]{Judge+Reardon+Cauzzi2012}
{Judge}, P.~G., {Reardon}, K., and {Cauzzi}, G.: 2012,
\newblock {\em Astrophys.\ J.\ Lett.\/} {\bf 755}, L11

\bibitem[\protect\astroncite{{Judge} {\em
  et~al.}}{2011}]{Judge+Tritschler+Low2011}
{Judge}, P.~G., {Tritschler}, A., and {Low}, B.~C.: 2011,
\newblock {\em \apjl\/} {\bf 730}, L4

\bibitem[\protect\astroncite{{Klimchuk}}{2012}]{Klimchuk2012}
{Klimchuk}, J.~A.: 2012,
\newblock {\em Journal of Geophysical Research (Space Physics)\/} {\bf 117},
  12102

\bibitem[\protect\astroncite{{Kuhn}}{1970}]{Kuhn1970}
{Kuhn}, T.~S.: 1970,
\newblock {\em {The structure of scientific revolutions}\/}

\bibitem[\protect\astroncite{{Langangen} {\em
  et~al.}}{2008}]{Langangen+others2008}
{Langangen}, {\O}., {De Pontieu}, B., {Carlsson}, M., {Hansteen}, V.~H.,
  {Cauzzi}, G., and {Reardon}, K.: 2008,
\newblock {\em Astrophys.\ J.\ Lett.\/} {\bf 679}, L167

\bibitem[\protect\astroncite{{Lemen} {\em et~al.}}{2012}]{Lemen+others2012}
{Lemen}, J.~R., and 46 co-authors: 2012,
\newblock {\em Solar Phys.\/} {\bf 275}, 17

\bibitem[\protect\astroncite{{L{\"o}fdahl}}{2002}]{Lofdahl2002}
{L{\"o}fdahl}, M.~G.: 2002,
\newblock in {P.~J.~Bones, M.~A.~Fiddy, \& R.~P.~Millane} (Ed.), {\em Society
  of Photo-Optical Instrumentation Engineers (SPIE) Conference Series\/}, Vol.
  4792 of {\em Society of Photo-Optical Instrumentation Engineers (SPIE)
  Conference Series\/}, p.~146

\bibitem[\protect\astroncite{{L{\'o}pez Ariste} and
  {Casini}}{2005}]{Lopez-Ariste+Casini2005}
{L{\'o}pez Ariste}, A. and {Casini}, R.: 2005,
\newblock {\em Astron.\ Astrophys.\/} {\bf 436}, 325

\bibitem[\protect\astroncite{{Miyamoto}}{1949}]{Miyamoto1949a}
{Miyamoto}, S.: 1949,
\newblock {\em Publ.\ Astron.\ Soc.\ Japan\/} {\bf 1}, 14

\bibitem[\protect\astroncite{{Nakariakov} and
  {Verwichte}}{2005}]{Nakariakov+Verwichte2005}
{Nakariakov}, V.~M. and {Verwichte}, E.: 2005,
\newblock {\em Living Reviews in Solar Physics\/} {\bf 2}, 3

\bibitem[\protect\astroncite{{Nistic{\`o}} {\em
  et~al.}}{2009}]{Nistico+others2009}
{Nistic{\`o}}, G., {Bothmer}, V., {Patsourakos}, S., and {Zimbardo}, G.: 2009,
\newblock {\em Solar Phys.\/} {\bf 259}, 87

\bibitem[\protect\astroncite{{Okamoto} and {De
  Pontieu}}{2011}]{Okamoto+dePontieu2011}
{Okamoto}, T.~J. and {De Pontieu}, B.: 2011,
\newblock {\em Astrophys.\ J.\/} {\bf 736}, L24

\bibitem[\protect\astroncite{Parker}{1994}]{Parker1994}
Parker, E.~N.: 1994,
\newblock {\em Spontaneous Current Sheets in Magnetic Fields with Application
  to Stellar X-Rays\/},
\newblock International Series on Astronomy and Astrophyics, Oxford University
  Press, Oxford

\bibitem[\protect\astroncite{{Patsourakos} {\em
  et~al.}}{2013}]{Patsourakos+Klimchuk+Young2013}
{Patsourakos}, S., {Klimchuk}, J., and {Young}, P.: 2013,
\newblock {\em ArXiv e-prints\/}

\bibitem[\protect\astroncite{{Pereira} {\em
  et~al.}}{2012}]{Pereira+dePontieu+Carlsson2012}
{Pereira}, T.~M.~D., {De Pontieu}, B., and {Carlsson}, M.: 2012,
\newblock {\em Astrophys.\ J.\/} {\bf 759}, 18

\bibitem[\protect\astroncite{{Pereira} {\em
  et~al.}}{2013}]{Pereira+dePontieu+Carlsson2013}
{Pereira}, T.~M.~D., {De Pontieu}, B., and {Carlsson}, M.: 2013,
\newblock {\em Astrophys.\ J.\/} {\bf 764}, 69

\bibitem[\protect\astroncite{Pietarila and Judge}{2004}]{Pietarila+Judge2004}
Pietarila, A. and Judge, P.~G.: 2004,
\newblock {\em Astrophys.\ J.\/} {\bf 606}, 1239

\bibitem[\protect\astroncite{Pneuman and Kopp}{1978}]{Pneuman+Kopp1978}
Pneuman, G. and Kopp, R.: 1978,
\newblock {\em Sol. Phys.\/} {\bf 57}, 49

\bibitem[\protect\astroncite{{Popper}}{1972}]{Popper1972}
{Popper}, K.~R.: 1972,
\newblock {\em {Conjectures and Refutations. The Growth of Scientific
  Knowledge}\/},
\newblock Routledge, London

\bibitem[\protect\astroncite{{Priest} {\em et~al.}}{2002}]{Priest+others2002}
{Priest}, E.~R., {Heyvaerts}, J.~F., and {Title}, A.~M.: 2002,
\newblock {\em Astrophys.\ J.\/} {\bf 576}, 533

\bibitem[\protect\astroncite{Roberts}{1945}]{Roberts1945}
Roberts, W.~O.: 1945,
\newblock {\em Astrophys.\ J.\/} {\bf 101}, 136

\bibitem[\protect\astroncite{{Rouppe van der Voort} {\em
  et~al.}}{2009}]{Rouppe+others2009}
{Rouppe van der Voort}, L., {Leenaarts}, J., {de Pontieu}, B., {Carlsson}, M.,
  and {Vissers}, G.: 2009,
\newblock {\em Astrophys.\ J.\/} {\bf 705}, 272

\bibitem[\protect\astroncite{{Rutten}}{2006}]{Rutten2006}
{Rutten}, R.~J.: 2006,
\newblock in {J.~Leibacher, R.~F.~Stein, \& H.~Uitenbroek} (Ed.), {\em Solar
  MHD Theory and Observations: A High Spatial Resolution Perspective\/}, Vol.
  354,  276

\bibitem[\protect\astroncite{{Rutten}}{2007}]{Rutten2007}
{Rutten}, R.~J.: 2007,
\newblock in P. {Heinzel}, I. {Dorotovi{\v c}}, and R.~J. {Rutten} (Eds.), {\em
  The Physics of Chromospheric Plasmas\/}, Vol. 368 of {\em Astronomical
  Society of the Pacific Conference Series\/}, ~27

\bibitem[\protect\astroncite{{Sasso} {\em et~al.}}{2012}]{Sasso+others2012}
{Sasso}, C., {Andretta}, V., {Spadaro}, D., and {Susino}, R.: 2012,
\newblock {\em Astron.\ Astrophys.\/} {\bf 537}, A150

\bibitem[\protect\astroncite{{Sekse} {\em et~al.}}{2012}]{Sekse+others2012}
{Sekse}, D.~H., {Rouppe van der Voort}, L., and {De Pontieu}, B.: 2012,
\newblock {\em Astrophys.\ J.\/} {\bf 752}, 108

\bibitem[\protect\astroncite{{Sekse} {\em
  et~al.}}{2013a}]{Sekse+Rouppe+dePontieu2013}
{Sekse}, D.~H., {Rouppe van der Voort}, L., and {De Pontieu}, B.: 2013a,
\newblock {\em Astrophys.\ J.\/} {\bf 764}, 164

\bibitem[\protect\astroncite{{Sekse} {\em et~al.}}{2013b}]{Sekse+others2013}
{Sekse}, D.~H., {Rouppe van der Voort}, L., {De Pontieu}, B., and {Scullion},
  E.: 2013b,
\newblock {\em Astrophys.\ J.\/} {\bf 769}, 44

\bibitem[\protect\astroncite{{Sterling}}{2000}]{Sterling2000}
{Sterling}, A.~C.: 2000,
\newblock {\em Solar Phys.\/} {\bf 196}, 79

\bibitem[\protect\astroncite{{Thomas}}{1948}]{Thomas1948a}
{Thomas}, R.~N.: 1948,
\newblock {\em Astrophys.\ J.\/} {\bf 108}, 130

\bibitem[\protect\astroncite{{Tripathi} and
  {Klimchuk}}{2013}]{Tripathi+Klimchuk2013}
{Tripathi}, D. and {Klimchuk}, J.~A.: 2013,
\newblock {\em Astrophys.\ J.\/} {\bf 779}, 1

\bibitem[\protect\astroncite{{van Ballegooijen} {\em
  et~al.}}{1998}]{vanBallegooijen+others1998}
{van Ballegooijen}, A.~A., {Nisenson}, P., {Noyes}, R.~W., {L{\"o}fdahl},
  M.~G., {Stein}, R.~F., {Nordlund}, {\AA}., and {Krishnakumar}, V.: 1998,
\newblock {\em Astrophys.\ J.\/} {\bf 509}, 435

\bibitem[\protect\astroncite{{W{\"o}ger} {\em
  et~al.}}{2008}]{Woeger+vonderLuhe+Reardon2008}
{W{\"o}ger}, F., {von der L{\"u}he}, O., and {Reardon}, K.: 2008,
\newblock {\em Astron.\ Astrophys.\/} {\bf 488}, 375

\bibitem[\protect\astroncite{{Zhang} {\em et~al.}}{2012}]{Zhang+others2012}
{Zhang}, Y.~Z., {Shibata}, K., {Wang}, J.~X., {Mao}, X.~J., {Matsumoto}, T.,
  {Liu}, Y., and {Su}, J.~T.: 2012,
\newblock {\em Astrophys.\ J.\/} {\bf 750}, 16

\end{thebibliography}
\end{document}